\documentclass[12pt,preprint]{aastex}

\begin{document}

\shorttitle{Damping of turbulence in partially ionized gas}
\shortauthors{Falceta-Gon\c calves, Lazarian \& Houde}

\title{Damping of MHD turbulence in partially ionized gas and the observed difference of 
velocities of neutrals and ions}
\author{D. Falceta-Gon\c calves\altaffilmark{1}, A. Lazarian\altaffilmark{2} \& M. 
Houde\altaffilmark{3}}
\altaffiltext{1}{N\' ucleo de Astrof\' isica Te\' orica, Universidade Cruzeiro do
Sul - Rua Galv\~ ao Bueno 868, CEP 01506-000, S\~ao Paulo, Brazil \\ 
email: diego.goncalves@cruzeirodosul.edu.br}
\altaffiltext{2}{Astronomy Department, University of Wisconsin,
 Madison, 475 N. Charter St., WI 53711, USA}
\altaffiltext{3}{Department of Physics and Astronomy, the University of Western Ontario, 
London, Ontario, Canada, N6A 3K7}

\begin{abstract}

Theoretical and observational studies on the turbulence of
the interstellar medium developed fast in the past decades. The 
theory of supersonic magnetized turbulence, 
as well as the understanding of projection effects of observed quantities, are 
still in progress. In this work we explore 
the characterization of the turbulent cascade and its damping from
observational spectral line profiles. We address the difference of ion and 
neutral velocities by clarifying the nature of the turbulence damping in 
the partially ionized. We provide theoretical arguments in favor of the explanation of 
the larger Doppler broadening of lines arising from neutral species compared to ions as arising from the turbulence 
damping of ions at larger scales. Also, we compute a number of MHD numerical
simulations for different turbulent regimes and explicit turbulent damping, and compare
both the 3-dimensional distributions of velocity and the synthetic line profile 
distributions.
From the numerical simulations, we place constraints 
on the precision with which one can measure the 3D dispersion  depending on the 
turbulence sonic Mach number. We show 
that no universal correspondence between the 3D velocity dispersions measured in the 
turbulent volume and minima of the 2D velocity dispersions available through observations 
exist. For instance, for 
subsonic turbulence the correspondence is poor at scales much smaller than the turbulence 
injection scale, while for supersonic turbulence the correspondence is poor for the 
scales comparable with the injection scale. We provide a physical explanation of the existence of such a 2D-3D correspondence 
and discuss the uncertainties in evaluating the damping scale of ions that can be 
obtained from observations. However, we show that the statistics 
of velocity dispersion from observed line profiles can provide the spectral index and the 
energy transfer rate of turbulence. Also, comparing two similar simulations with 
different viscous coefficients it was possible to constrain the turbulent cut-off scale. 
This may especially prove useful since it is believed 
that ambipolar diffusion may be one of the dominant dissipative mechanism in star-forming 
regions. In this case, the determination of the ambipolar diffusion scale may be used as 
a complementary method for the determination of magnetic field intensity in collapsing 
cores.  We discuss 
the implications of our findings in terms of a new approach to magnetic field measurement 
proposed by Li \& Houde (2008).      
\end{abstract}

\section{Introduction}

The interstellar medium (ISM) is known to be composed by a multi-phase, turbulent and 
magnetized gas (see Brunt \& Heyer 2002, Elmegreen \& Scalo 2004, Crutcher 2004, McKee \& 
Ostriker 2007). 
However, the relative importance of turbulence and the magnetic field in the ISM dynamics 
and in the formation of structures is still a matter of debate. More specifically, typical molecular clouds 
present densities in the range of $10^2 - 10^5$ cm$^{-3}$, sizes $L \sim 0.1 - 100$pc, 
temperature $T \sim 10 - 20$ K, and lifetimes that are larger than the Jeans gravitational collapse timescale. 
The role of the magnetic field in preventing the collapse is hotly debated in the 
literature (see Fiedge \& Pudritz 2000, Falceta-Gon\c calves, de Juli \& Jatenco-Pereira 2003, MacLow \& Klessen 2004). 
Magnetic field can be removed from clouds in the presence of ambipolar diffusion arising 
from the differential drift of neutrals and ions (Mestel \& Spitzer 1986, Shu 1983) and reconnection diffusion which 
arises from fast magnetic reconnection of turbulent magnetic field (Lazarian 2005). 
Nevertheless, 
the role of magnetic fields in the dynamics of ISM is difficult to underestimate. 

We feel that a lot of the unresolved issues in the theory of star formation are in part due to the fact that the 
amount of information on magnetic fields obtainable through presently used techniques is 
very limited. For molecular cloud the major ways of obtaining information about magnetic fields amount to Zeeman 
broadening of spectral lines, which provides measures of the field strength along the 
line of sight (see Crutcher 1999) and the Chandrashekar-Fermi (CF) method, which uses the statistics of polarization 
vectors to provide the amplitude of the plane of the sky component of the field (see 
Hildebrand 2000). However, since Zeeman measurements are restricted to rather strong magnetic fields (due to current 
observational sensitity) and therefore the measurements are restricted to dense clouds 
and new measurements require a lot of observational time. At the same time,  
the CF method relies on all grains being perfectly aligned, which is known not to be the case in molecular clouds 
(see Lazarian 2007 for a review). The CR technique is also known to systematically 
overestimate the field intensity (Houde et al. 2009, Hildebrand et al. 2009), and to poorly map the magnetic field 
topology for super-Alfvenic turbulence (see Falceta-Gon\c calves, Lazarian \& Kowal 2008).

The difficulty of the traditional techniques call for new approaches in measuring astrophysical magnetic fields. 
Recently, a number of such techniques has been proposed. For instance, Yan \& Lazarian 
(2006, 2007, 2008) discussed using the radiative alignment of atoms and ions having fine or hyperfine split of the 
ground of metastable levels. The technique is based on the successful alignment of atoms 
in the laboratory conditions, but it requires environments where radiative pumping dominates the collisional 
de-excitation of the levels.    

Another new approach which we dwell upon in this paper is based on the comparison of 
the ion-neutral spectral lines. Houde et al. (2000a, 2000b) identified the differences of the width of the lines of 
neutral atoms and ions as arising from their differential interaction with magnetic 
fields. It was assumed that because ions are forced into gyromagnetic motions about magnetic field lines that their 
spectral line profiles would thus reveal the imprint of the magnetic field on their 
dynamics.

In particular, as observations of HCN and HCO$^{+}$ in molecular clouds revealed significantly and systematically 
narrower ion lines, Houde et al. (2000a) proposed a simple explanation for these 
observations. The model was solely based on the strong Lorentz interaction between the ion and the magnetic field 
lines, but also required the presence of turbulent motions in the gas. More precisely, it 
was found that the observations of the narrower HCO$^{+}$ lines when compared to that from the coexistent HCN species 
could potentially be explained if neutral particles stream pass magnetic field lines with 
the entrained ions. Such a picture could be a particular manifestation of the ambipolar diffusion phenomenon.  

Although this model was successful in explaining the differences between the velocities of ions and neutrals, the 
quantitative description of the model of drift was oversimplified. For example, it was 
neither possible to infer anything about the strength of the magnetic field nor was the "amount" of ambipolar diffusion, 
which is at the root of the observable effect described by the model, quantifiable in any 
obvious manner. The main reason for these shortcomings resides in the way that turbulence and its interplay with the 
magnetic field were treated in the analysis of Houde et al. (2000a); a more complete and 
powerful model was required.	
The next step in the study of magnetized turbulence and ambipolar diffusion through the comparison of the coexistent 
ion/neutral spectral lines was taken by Li \& Houde (2008) where a model of turbulence 
damping in partially ionized gas was employed. Assumption that the main damping mechanism is associated with ambipolar 
diffusion, they deduced proposed a way for evaluating the strength of the 
plane-of-the-sky component of the magnetic field in molecular clouds.

The main idea is that, in a magnetically dominated scenario, cloud collapse and magnetic 
energy removal may 
be accelerated due to ambipolar diffusion of ions and neutral particles. Although as gravity becomes dominant the collapsing 
cloud continuously drags material to its core including the ions, which are frozen to the 
magnetic field lines, magnetic pressure slows down their infall, 
but not that of the neutrals. At this stage most of the matter, in neutral phase, continues to decouple from the 
ionic fluid and the field lines leading to the diffusion of 
magnetic energy and the collapse may develop further. This ion-neutral drift, excited by 
the ambipolar diffusion, is also responsible for damping the ion turbulent motions. The 
increase in the net viscosity of the flow provides a cut-off in the turbulent cells with 
turnover timescales lower than the period of collisions $\tau_{i,n}$ (see Lazarian, 
Vishniac \& Cho [2004] for detailed review).

Since the turbulent cascade 
is dramatically changed by the decoupling of the ion and neutral fluids the 
observed velocity dispersion could reveal much of the physics of collapse length 
scales. The interpretation and reliability of this technique, however, still need to be corroborated with more 
detailed theoretical analysis, as well as numerical simulations of 
magnetized turbulence.

For the past decade, because of its complicated, fully non-linear 
and time-dependent nature, magnetized turbulence has been mostly studied by numerical 
simulations (see Ostriker, Stone \& Gammie 2001, Cho \& Lazarian 2005, Kowal, Lazarian \& 
Beresniak 2007). Simulations can uniquely provide the three-dimensional structure for the 
density, velocity and magnetic fields, as well as two dimensional maps that can be 
compared to observations (e.g. column density, line profiles, polarization maps). 
Therefore, direct comparison of observed and synthetic maps may help reveal the magnetic 
topology and velocity structure.

In this paper we re-examine the assumptions made in this model and test some of these assumptions using the MHD 
numerical simulations. In particular, we provide a number of numerical simulations of MHD 
turbulent flows, with different sonic and Alfvenic Mach numbers. In \S2, we describe the NIDR technique for the 
determination of the damping 
scales and magnetic field intensity from dispersion of velocities and the main 
theoretical aspects of turbulence in partially ionized gases. In \S3, we describe 
the numerical simulations and present the results and the statistical analysis of the data. In \S4, we discuss 
the systematic errors intrinsic to the procedures involved, followed by the discussion of 
the results and summary, in \S5.

\section{Turbulence in Partially Ionized Gas}

\subsection{Challenge of interstellar turbulence}

In 1941, Kolmogorov proposed the well-known theory for energy cascade in incompressible 
fluids. Under Kolmogorov's approximation, turbulence evolves from the largest to smaller 
scales, up to the dissipation scales, as follows. Within the so-called inertial range, 
i.e the range of scales large enough for dissipation to be negligible but still smaller 
than the injection scales, the energy spectrum may be well described by, 

\begin{equation}
P(k) \sim \dot{\epsilon}^{\alpha-1} k^{-\alpha},
\end{equation}

\noindent
where $\dot{\epsilon}$ is the energy transfer rate between scales and $\alpha \sim 5/3$. 
In this approximation, within the inertial range, the energy transfer rate is assumed to 
be constant for all scales. Therefore, integrating Eq. 3 over $k$ for $\alpha = 5/3$, we 
obtain,

\begin{equation}
\sigma^2 (k) \propto \left( \frac{3 \dot{\epsilon}^{2/3}}{2} \right) k^{-2/3}.
\end{equation}

However, reality is far more complicated. First, the ISM is threaded by magnetic fields, 
which may be strong enough to play a role on the dynamics of eddies and change the 
scaling relations. Second, observations suggest that the ISM is, at large scales, is 
highly compressible. Third, many phases of the ISM (see Draine \& Lazarian 1999 for 
typical parameters) are partially ionized. 

Attempts to include magnetic fields in the picture of turbulence include works by Iroshnikov (1964) and 
Kraichnam (1965), which were done assuming that magnetized turbulence stays isotropic. 
Later studies proved that magnetic field introduces anisotropy into turbulence 
(Shebalin, Matthaeus, \& Montgomery 1983, Higdon 1984, Zank \& 
Matthaeus 1992, see also book by Biskamp 2003).  

Goldreich \& Sridhar (1995, henceforth GS95) proposed a model for magnetic incompressible 
turbulence\footnote{In the original treatment of GS95 the description of turbulence is limited to a situation 
of the velocity of injection at the injection scale $V_L$ being equal to $V_A$. The 
generalization of the scalings when $V_L<V_A$ can be found in Lazarian \& Vishniac (1999). The generalization for 
the $V_L>V_A$ is also straightforward (see Lazarian 2006).} based on the anisotropies in 
scaling relations, as eddies would evolve 
differently in directions parallel and perpendicular to the field lines: 
$V_A/\Lambda_{\parallel} \sim v_{\lambda}/\lambda$, where $\Lambda_{\parallel}$ is the parallel scale of the eddy and $\lambda$ is its perpendicular scale. These scales are measured in respect to the {\it local}\footnote{The latter issue does not formally allow to describe turbulence in the Fourier space, as the latter calls for the description in respect to the global magnetic field.} magnetic field. Combining this to the assumption of 
self-similarity in energy transfer rate, we get a Kolmogorov-like spectrum for 
perpendicular motions with $\alpha =5/3$ and, most importantly, the anisotropy in the 
eddies scales as $\Lambda_\parallel \propto \lambda^{2/3}$. 

In spite of the intensive recent work on the incompressible turbulence (see Boldyrev 2005, 2006, Beresnyak \& Lazarian 2006, 2009), we feel that the GS95 is the model that can guide us in the research in the absence of a better alternative. The generalization of the GS95 for compressible motions are available (Lithwick \& Goldreich 2001, Cho \& Lazarian 2002, 2003) and they consider scalings of the fast and slow MHD modes.

\subsection{Turbulence damping in partially ionized gas}

{\bf

As stated before, the turbulent cascade is expected to develop down to scales where 
dissipation processes become dominant. The dissipation scales are associated to the viscous damping, which is responsible for the transfer of kinetic into thermal energy of any eddy smaller than the viscous cutoff scale. 

In the ISM, e.g. in cold clouds, the gas is partially ionized and the coupling 
between neutrals, ions and magnetic fields gives rise to interesting processes. As far as 
damping is concerned one of the most interesting is the energy dissipation as the motions of ions and 
neutral particles decouple. While the issue of turbulence dissipation has been discussed 
extensively in the literature (see Minter \& Spangler 1997), a generalization of the GS95 model of turbulence 
for the case of the partially ionized gas was presented in Lazarian, Vishniac \& 
Cho (2004) (LVC04). 
In their model, if the eddy turnover time ($\tau$) gets of order of the ion-neutral 
collision rate ($t_{in}^{-1}$) two fluids are strongly coupled. In this situation a 
cascade cut-off is present.

In a strongly coupled fluid, using the scaling relation for the inertial range $v_{\rm 
damp} \sim U_{\rm inj}(L_{\rm inj}^{-1} l_{\rm damp})^{1/3}$, the damping scale is 
given by (LVC04),

\begin{equation}
l_{\rm damp} \sim \lambda_{\rm mfp} \left( \frac{c_n}{v_l} \right) \left( 
\frac{V_A}{v_l} \right)^{1/3} f_n,
\end{equation}

\noindent
where $f_n$ is the neutral fraction, $\lambda_{\rm mfp}$ and 
$c_n$ the mean free path and sound speed for the neutrals, respectively, $V_A$ is the 
Alfv\'en speed, and subscript ``inj" refers to the injection scale. Since the Alfvén 
speed depends on the magnetic field, $V_A = B (4 \pi \rho)^{-1/2}$, Eq.(3) is rewritten 
as:

\begin{equation}
B \sim \left( \frac{l_{\rm damp}}{\lambda_{\rm mfp}} \right)^{3} 
\left( \frac{v_l}{c_n} \right)^{3} (4 \pi \rho)^{1/2} f_n^{-3} v_l .
\end{equation}

Therefore, for a given molecular cloud, if the decoupling of ions and neutrals is the 
main process responsible for the ion turbulence damping Eq.(4) may be a 
complementary estimation of $B$. The main advantage of this method is 
that $B$ is the total magnetic field and not a component, parallel or perpendicular to 
the LOS, as respectively obtained from Zeeman or CF-method from polarization maps.}

\subsection{Approach by Li \& Houde 2008}

From the perspective of the turbulence above we can discuss the model adopted by 
Li \& Houde (2008) for their study. The authors considered that damping of ion motions happen earlier 
than those by neutrals at sufficiently small scales. At large 
scales, ions and neutrals are well coupled through flux freezing and their power spectra should be similar. 
At small scales the ion turbulence damps while the turbulence of neutral particles 
continues cascading to smaller scales. This difference may be detected in the velocity dispersions ($\sigma$) 
obtained from the integration of the velocity power spectrum over  the wavenumber $k$,

\begin{eqnarray}
\sigma^2 (k) & \propto & \int^{\infty}_{k} P(k')dk' \nonumber \\ 
& \sim & b k^{-n},
\end{eqnarray}

\noindent
considering $P(k)$ a power-law spectrum function. Since
the turbulence of ions is damped at the diffusion/dissipation scale ($L_{\rm D}$), while 
the turbulence of neutral particles may develop up to higher wavenumbers we may consider 
that the ions and neutral particles present the same distribution of velocities (well coupled) 
for $L>L_{\rm D}$. In this sense, the dispersion of neutral particles may be written as,

\begin{eqnarray}
\sigma^2_{\rm n} (k) & \propto & \int^{\infty}_{k} P(k')dk' \nonumber \\
 & = & \int^{k_{\rm D}}_{k} P(k')dk'+\int^{\infty}_{k_{\rm D}} P(k')dk' \nonumber \\ 
& \sim & \sigma^2_{\rm i} (k) + b k_{\rm D}^{-n},
\end{eqnarray}

\noindent
where $k_{\rm D} \approx L_{\rm D}^{-1}$. Eq.(5) may be directly compared to Eq. 
2. In this case, we would obtain $n \sim 2/3$ and $b$ is related to the energy transfer 
rate $\dot{\epsilon}$. Therefore, once the fitting parameters of Eq. 1 are obtained from 
the observational data, it is possible to obtain the cascading constants $\dot{\epsilon}$ 
and $\alpha$.

With the dispersion of velocities for both ions and neutrals at different scales $k$ it is possible to calculate 
the damping scale $k_{\rm D}$. From Eq.(5), the dispersion of neutral particles provides 
$b$ and $n$ constants and, by combining neutral and ion dispersions, it is possible to 
get $k_{\rm D}$ (Eq.6), i.e. the damping scale. 

Finally, as proposed by Li \& 
Houde (2008) in a different context, it is possible to to evaluate magnetic field 
strength by Eq.(4). We feel that the procedure of magnetic 
field study requires a separate discussion, due to its complexity, but 
in what follows we concentrate on the interesting facts of observational determining of 
the characteristics of turbulence and 
its damping that are employed in the technique by Li \& Houde (2008).

\subsection{Observational perspective}

Eqs. (1) and (2) are based 
on the dispersion of a three-dimensional velocity field, i.e. subvolumes with dimensions 
$k^{-3}$. 
Observational maps of line profiles, on the other hand, provide measurements of the 
velocity field integrated along the line of sight (LOS) within the area of the beam, i.e. 
a total volume of $k^{-2}k_{\rm min}^{-1}$ (with $k = 1/l$ and $k_{\rm min} = 1/L$, as 
$L$ represents the total depth of the structure observed - typically larger than $l$). 
Also, velocity dispersions are obtained from 
spectral line profiles, which are strongly dependent on the column density, i.e. the 
distribution of matter along the LOS. These factors make the comparison between observed 
lines and theoretical distribution of velocity fields a hard task. 

Fortunately, 3-dimensional numerical simulations of MHD turbulence may be useful in 
providing both the volumetric properties of the plasma parameters as well as their 
synthetic measurements projected along given lines of sight, which may be compared 
directly to observations, such as the spectral line dispersion. 
In this sense, based on the simulations of Ostriker et al. (2001), Li \& Houde 
(2008) stated that the actual dispersion of velocity is, approximately, the minimum 
value of the LOS dispersion, at each beam size $l$, obtained in 
a large sample of measurements. However, Ostriker et al. (2001) presented a single 
simulation, exclusively for supersonic and sub-alfvenic turbulent regime, with limited 
resolution ($256^3$). They also did not study increased viscosity, nor the correlation of 
minima of the synthetic dispersion and the turbulent regimes and the distribution of gas 
along the LOS.

In the following sections we will describe the details regarding the estimation of line 
dispersions, but now comparing it with a larger set of numerical simulations with
different turbulent regimes and with finer numerical resolution. The idea is to 
determine whether the technique is useful or not, and if there is any limitations with 
the different turbulent regimes. These tests are mandatory to ensure the applicability of 
the NIDR method to ISM observations.

\section{Numerical Simulations}

\begin{table*}
\begin{center}
\caption{Description of the simulations - $B$ is assumed in x-direction}
\begin{tabular}{cccc}
\hline\hline
Model & $M_S$ & $M_A$ & Description \\
\hline
1 & $0.7$ & $0.7$ & subsonic \& sub-Alfvenic\\
2 & $1.5$ & $0.7$ & supersonic \& sub-Alfvenic\\
3 & $7.0$ & $0.7$ & supersonic \& sub-Alfvenic\\
4 & $0.7$ & $7.0$ & subsonic \& super-Alfvenic\\
5 & $1.5$ & $7.0$ & supersonic \& super-Alfvenic\\
6 & $7.0$ & $7.0$ & supersonic \& super-Alfvenic\\
\hline\hline
\end{tabular}
\tablenotetext{}{sonic Mach number ($M_S = \left< v/c_s \right>$)}
\tablenotetext{}{Alfvenic Mach number ($M_A = \left< v/V_A \right>$)}
\end{center}
\end{table*}

In order to test the NIDR model, i.e to verify if the minimum dispersion of the 
velocity measured along the line of sight for a given beamsize $l \times l$ is 
aproximately the actual value calculated for a volume $l^3$, we used a total of 12 3-D MHD numerical 
simulations, with $512^3$ resolution, for 6 different turbulent regimes as described in 
Table 1, but repeated for viscous and inviscid models. 

The simulations were performed solving the set of ideal MHD isothermal equations, in 
conservative form, as follows:

\begin{equation}
\frac{\partial \rho}{\partial t} + \mathbf{\nabla} \cdot (\rho{\bf v}) = 0,
\end{equation}

\begin{equation}
\frac{\partial \rho {\bf v}}{\partial t} + \mathbf{\nabla} \cdot \left[ \rho{\bf v v} + 
\left( p+\frac{B^2}{8 \pi} \right) {\bf I} - \frac{1}{4 \pi}{\bf B B} \right] = {\bf f},
\end{equation}

\begin{equation}
\frac{\partial \mathbf{B}}{\partial t} - \mathbf{\nabla \times (v \times B)} = 0,
\end{equation}

\begin{equation}
\mathbf{\nabla \cdot B} = 0,
\end{equation}

\begin{equation}
p = c_s^2 \rho,
\end{equation}

\noindent
where $\rho$, ${\bf v}$ and $p$ are the plasma density, velocity and pressure, 
respectively, ${\bf B = \nabla \times A}$ is the magnetic field, ${\bf A}$ is 
the vector potential and ${\bf f} = {\bf f_{\rm turb}}+{\bf 
f_{\rm visc}}$ represents the external source terms, responsible for 
the turbulence injection and explicit viscosity. {\bf The code solves the set of MHD equations 
using a Godunov-type scheme, based on a second-order-accurate and the non-oscillatory spatial 
reconstruction (see Del Zanna et al. 2003). The shock-capture method is based on the Harten-Lax-van Leer 
(1983) Riemann solver. The magnetic divergence-free is assured by the use of a constrained transport method 
for the induction equation and the non-centered positioning of the magnetic field variables (see Londrillo 
\& Del Zanna 2000). The code has been extensively tested and successfully used in several works (Falceta-Gon\c 
calves, Lazarian \& Kowal 2008; Le{\~a}o et al. 2009; Burkhart et al. 2009; Kowal et al. 2009; Falceta-Gon\c 
calves et al. 2010). }

The turbulence is triggered by the injection of solenoidal perturbations in Fourier space of the velocity field.  Here, we solve the explicit 
viscous term as $\bf{f}_{\rm visc} = -\rho \nu \nabla ^2 {\bf v}$, where $\nu$ represents 
the viscous coefficient and is set arbitrarily to simulate the increased viscosity of the 
ionic flows due to the ambipolar diffusion. We run all the initial conditions given in 
Table 1 for 
both $\nu = 0$ and $\nu = 10^{-3}$, representing the neutral and ion particles fluids, 
respectively. Each simulations is initiated with an uniform density distribution, 
threaded by an uniform magnetic field. The simulations were run until the power spectrum 
is fully developed. The simulated box boundaries were set as periodic.

\begin{figure*}[tbh]
\centering
\includegraphics[scale=.50]{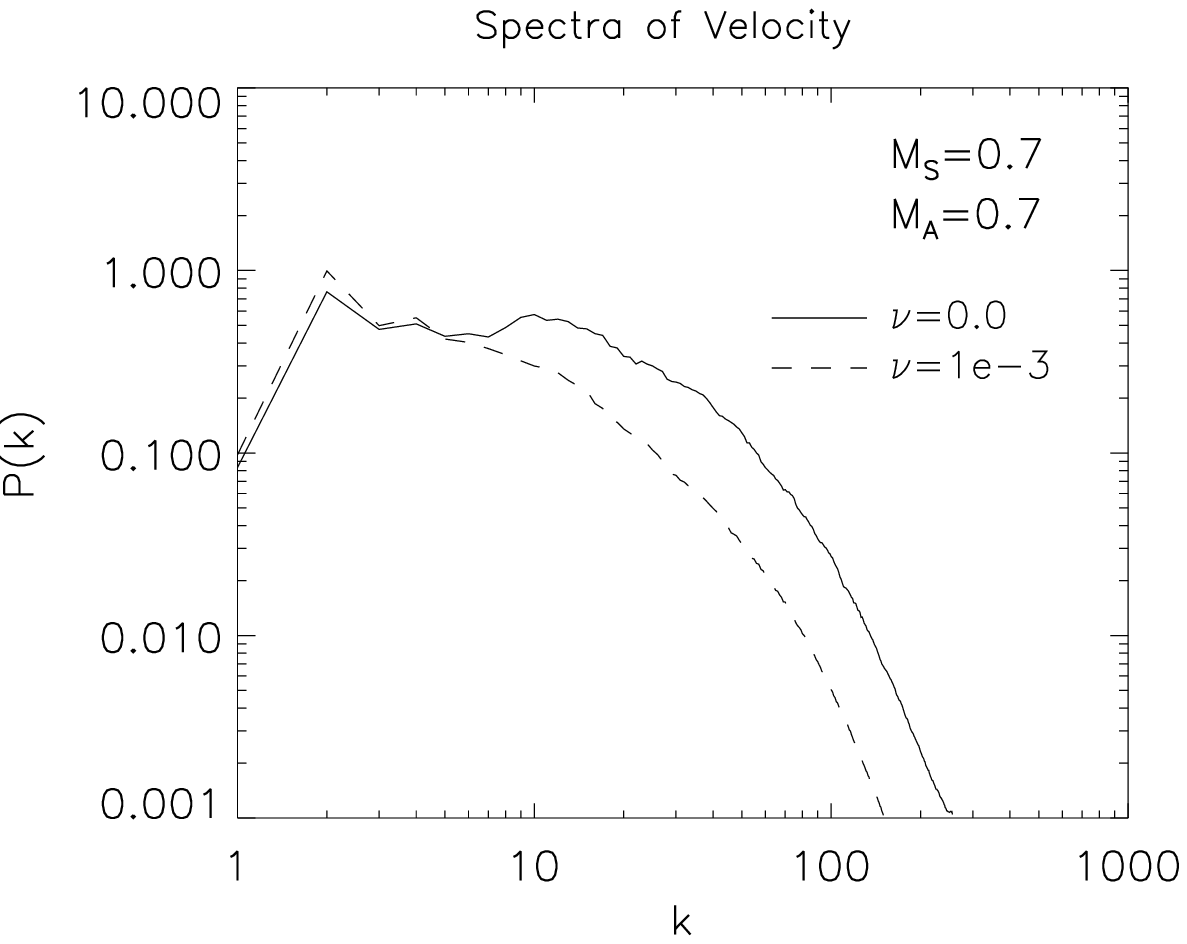}   
\includegraphics[scale=.50]{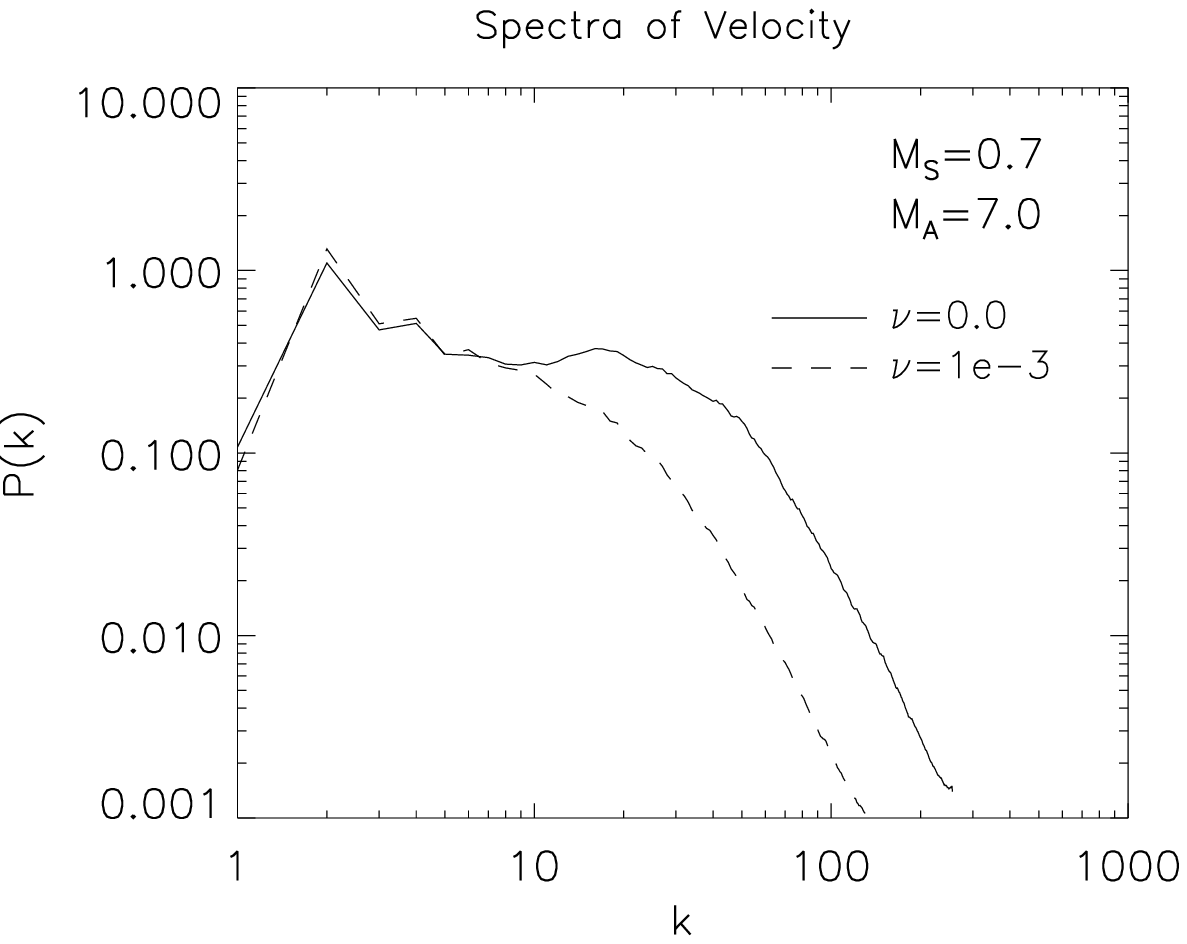}\\ 
\includegraphics[scale=.50]{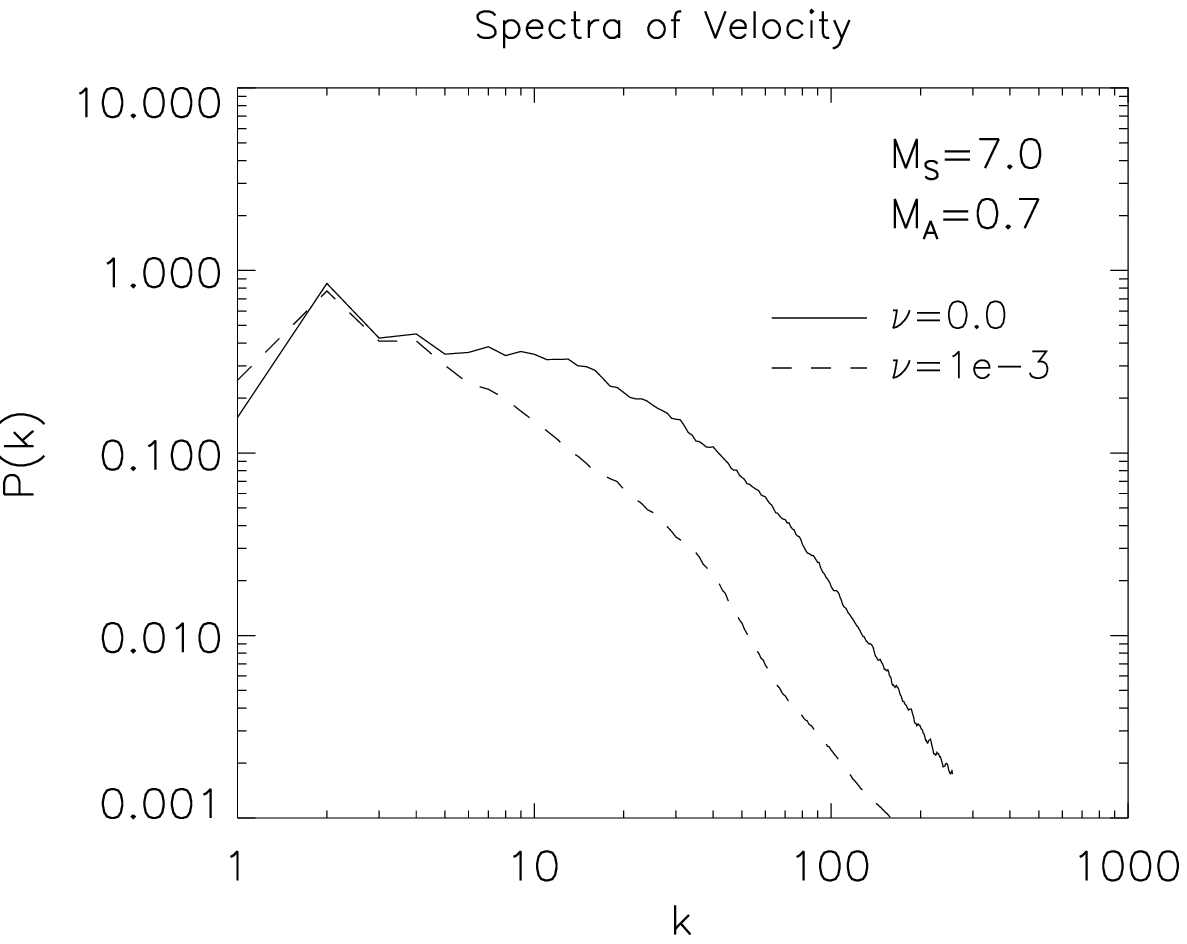}   
\includegraphics[scale=.50]{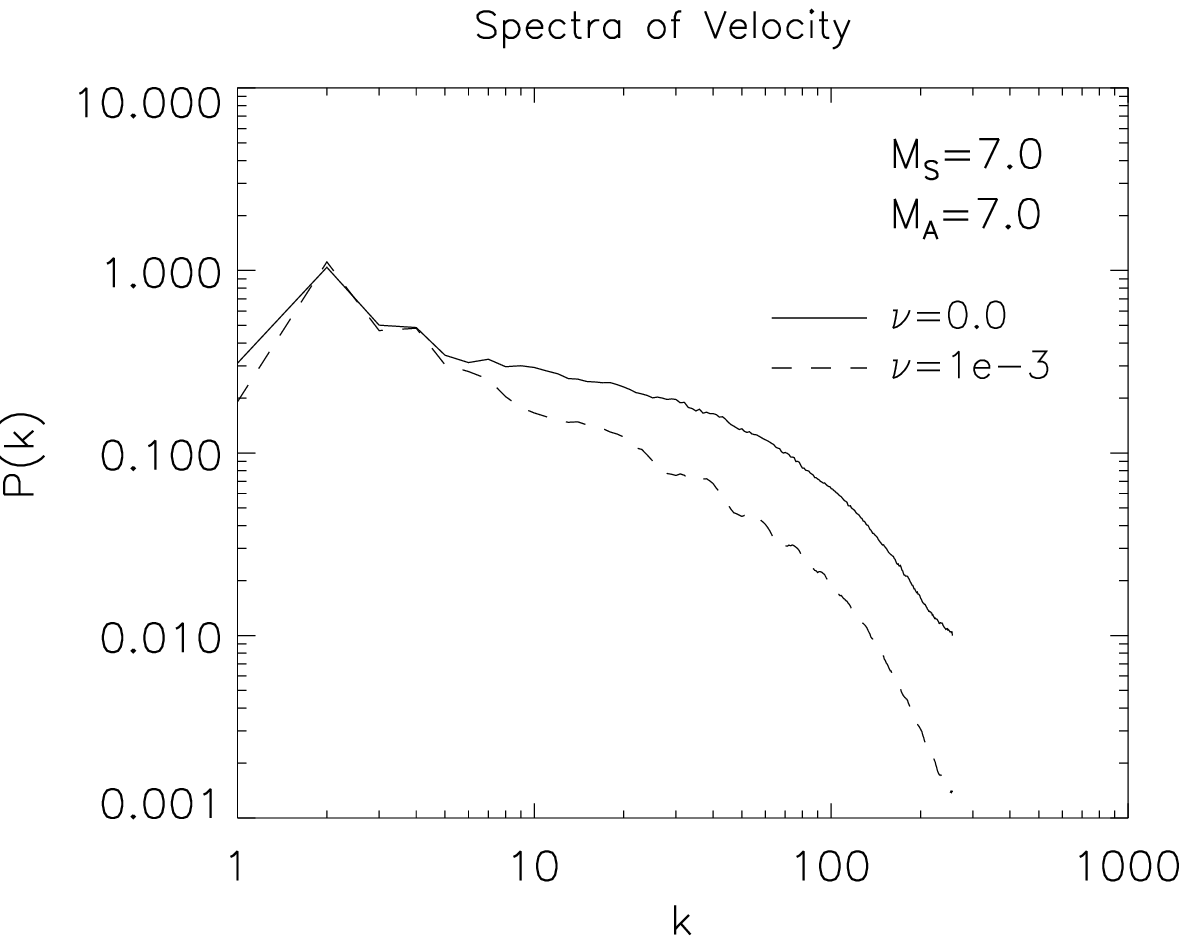}\\
\caption{Velocity power spectra of four of the models described in Table 1, separated by 
turbulent regime. The spectra are normalized with a Kolmogorov power function ($P \propto 
k^{-5/3}$). Solid lines represent the non-viscous fluid, and the dotted line the viscous 
fluid.}
\end{figure*}

In Fig. 1 we show the resulting velocity power spectra of four of our models, 
representing the four different turbulent regimes, i.e. (sub)supersonic and 
(sub)super-Alfv\'enic. 
Solid lines represent the non-viscous cases, and the dotted line the 
viscous cases. The spectra are 
normalized by a Kolmogorov power function $P \propto k^{-5/3}$. The inertial range 
of the scales is given by the horizontal part of the spectra. For the inviscid 
fluid, subsonic 
turbulence presents approximately flat spectra for $2 < k \leq 50$. Supersonic 
turbulence, on the other hand, shows steeper power spectra within this range. 
Actually, as shown from numerical simulations by Kritsuk et al. (2007) and Kowal \& 
Lazarian (2007), shocks in supersonic flows are responsible for the filamentation of 
structures and the increase in the energy flux cascade, resulting in a power spectrum 
slope $\sim -2.0$. For $k > 50$, the power spectra show a strong damping of the 
turbulence, resulting from the numerical viscosity. For the viscous fluid, the damped 
region is broadened ($k_{\rm cutoff} \sim 20$), due to the stronger 
viscosity.

\section{Relationship between 2D and 3D dispersion of velocities} 

\subsection{Comparing the synthetic to the 3-dimensional dispersion of velocities}

Theoretically, as given by Eq.(6), the difference between the two spectra for each run 
may be obtained from the observed dispersion of velocities.
The next step then is to obtain the dispersion of velocity, for different scales 
$l$, from our simulations. However, as explained previously, there are two different 
methods to obtain this parameter. One represents the actual dispersion, calculated within 
subvolumes $l^3$ of the computational box, while the second represents the 
observational measurements and is the dispersion of the 
velocity within the subvolume $l^2L$ (assuming the gas is optically thin), where $L$ 
is the total depth 
of the box. In order to match our calculations to observational measurements we will 
use the density weighted velocity $v^* = \rho v$ (see Esquivel \& Lazarian [2005]), which 
characterizes the line emission intensity proportional to the local density.

\begin{figure*}[tbh]
\centering
\includegraphics[scale=.50]{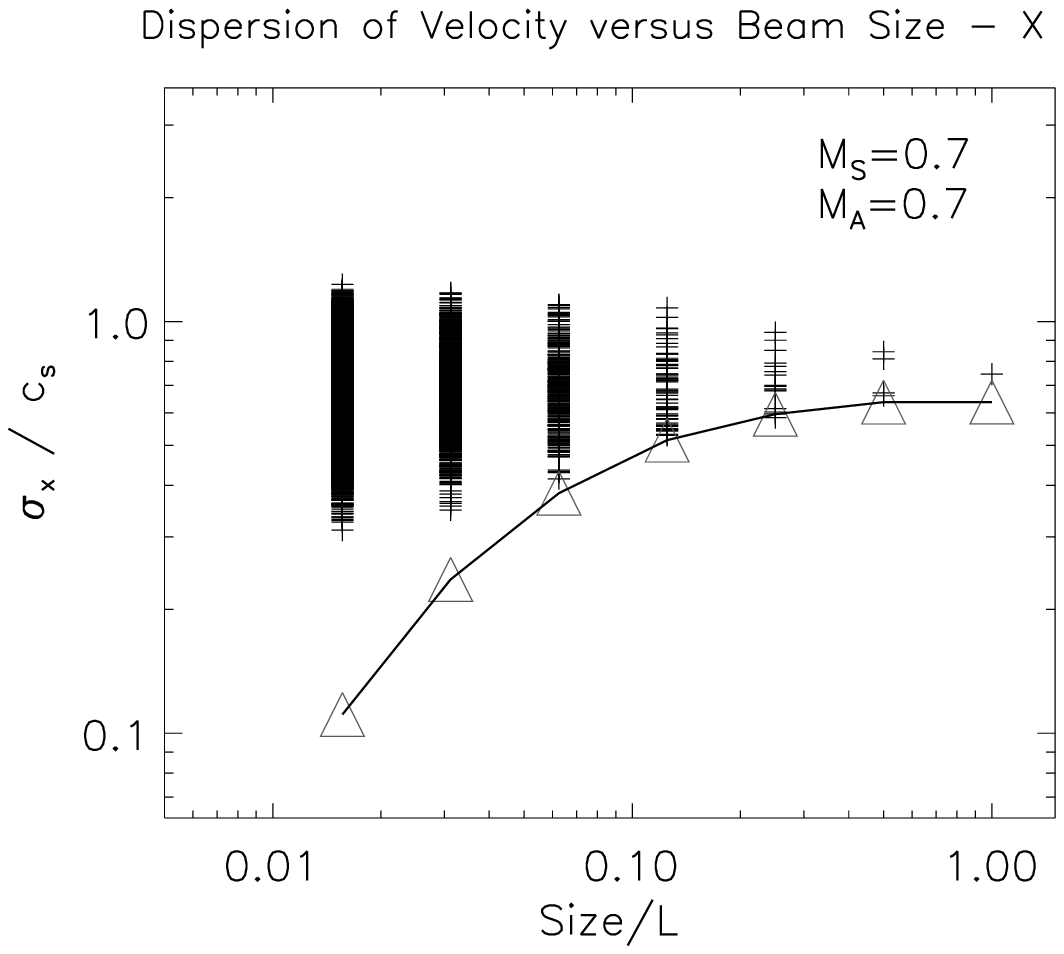}   
\includegraphics[scale=.50]{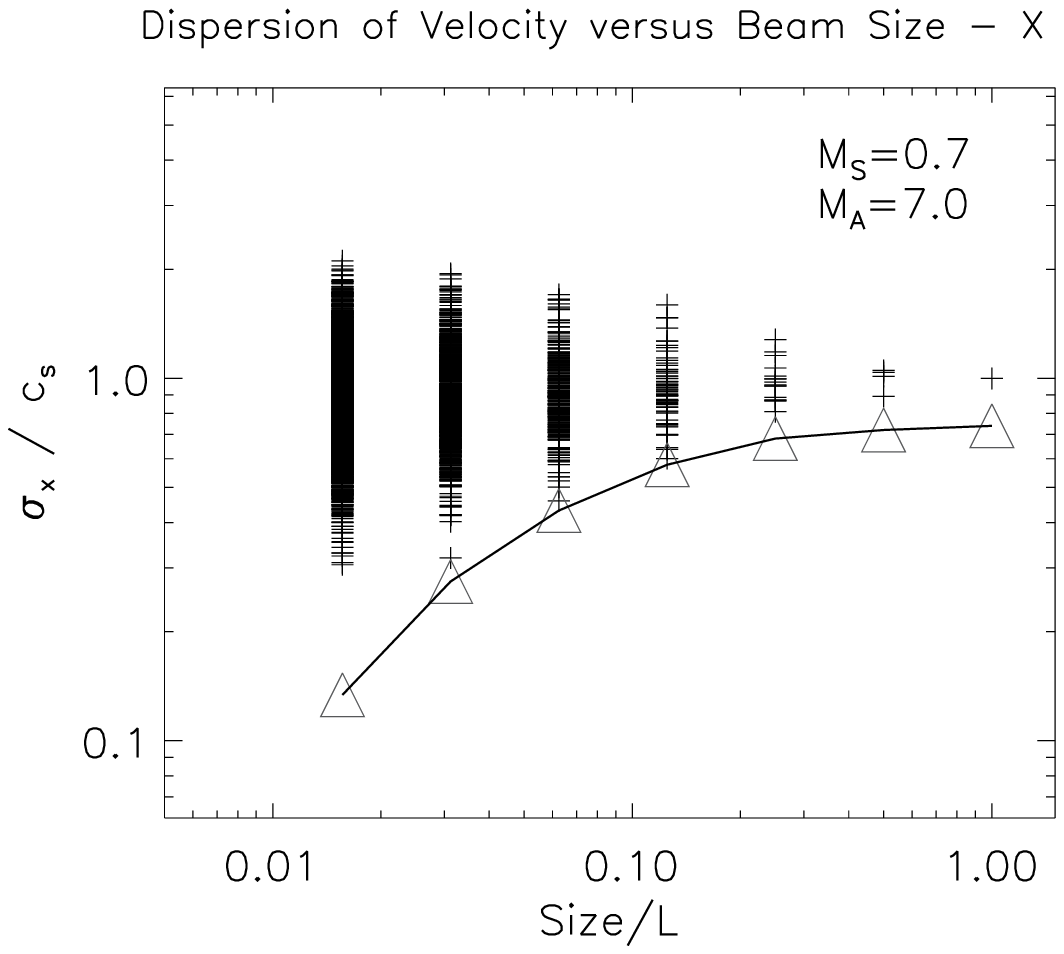}\\
\includegraphics[scale=.50]{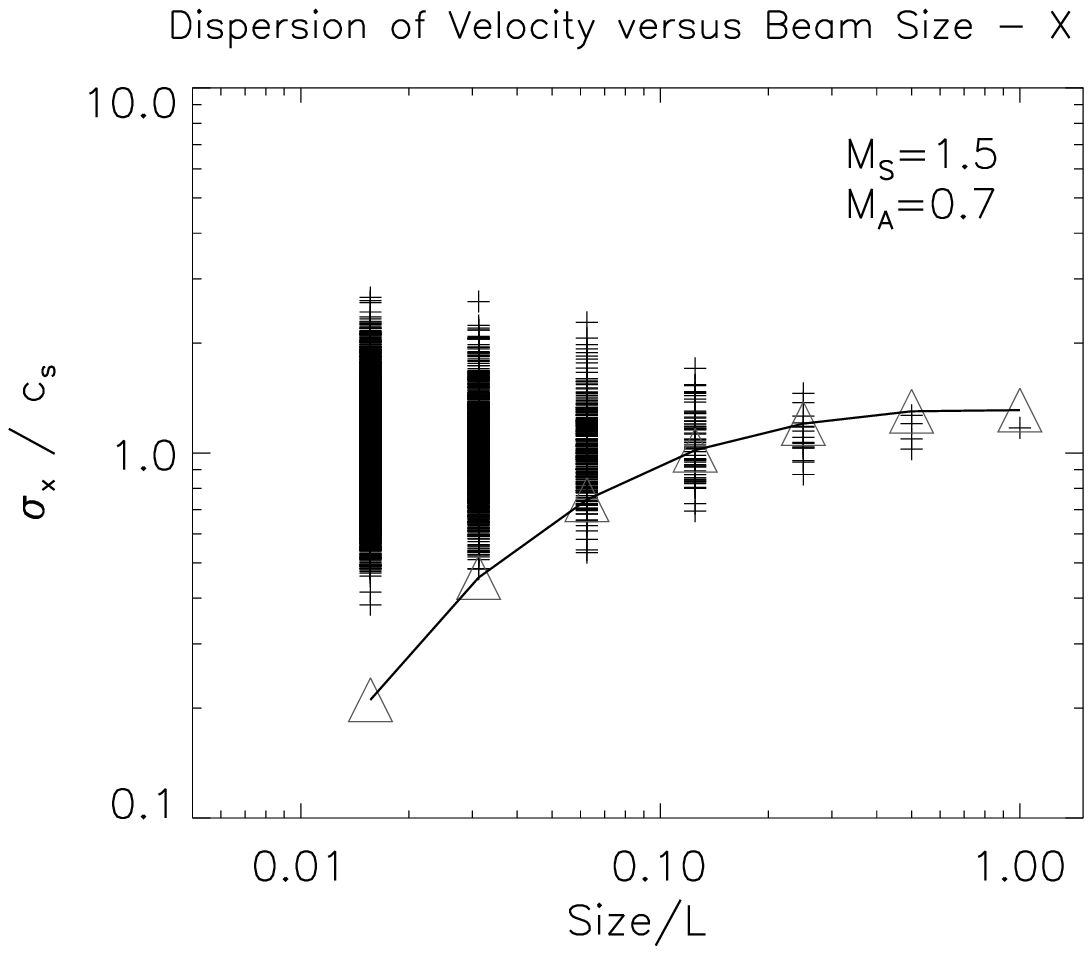}   
\includegraphics[scale=.50]{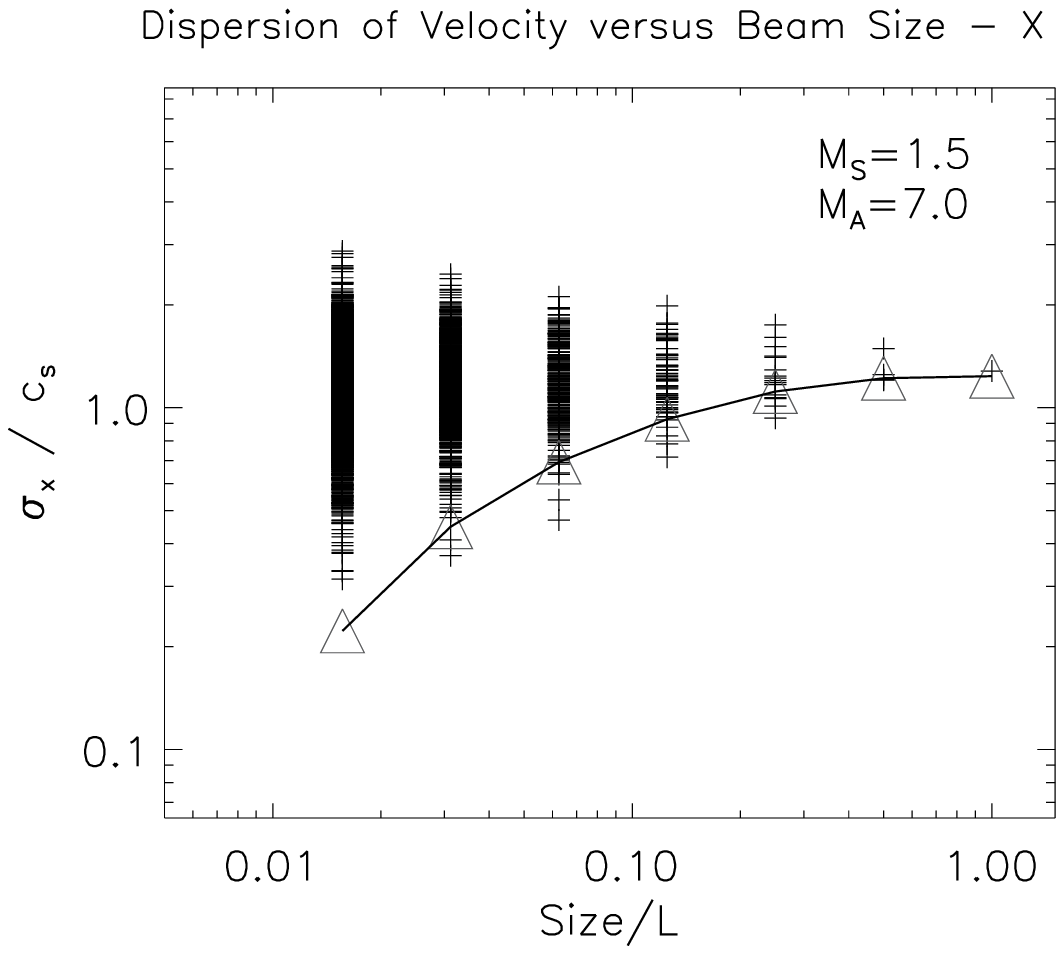}\\   
\includegraphics[scale=.50]{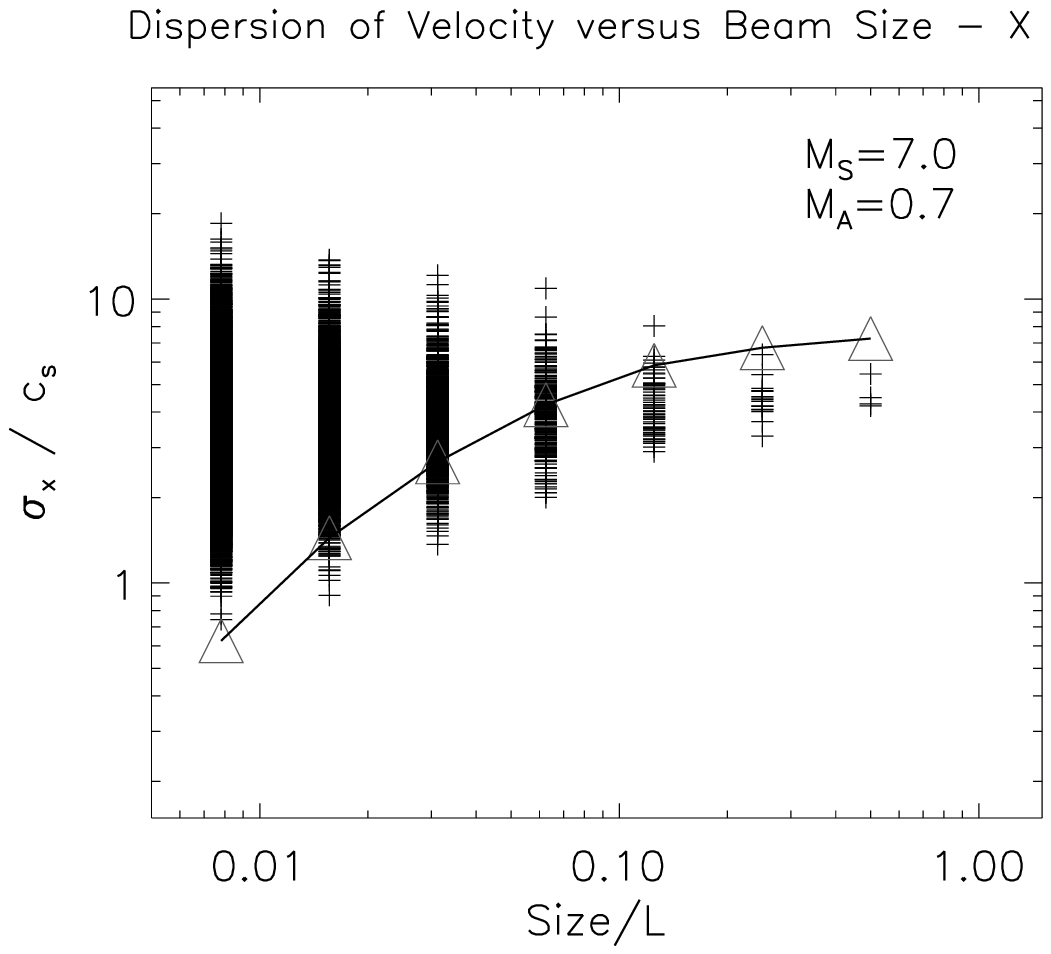}   
\includegraphics[scale=.50]{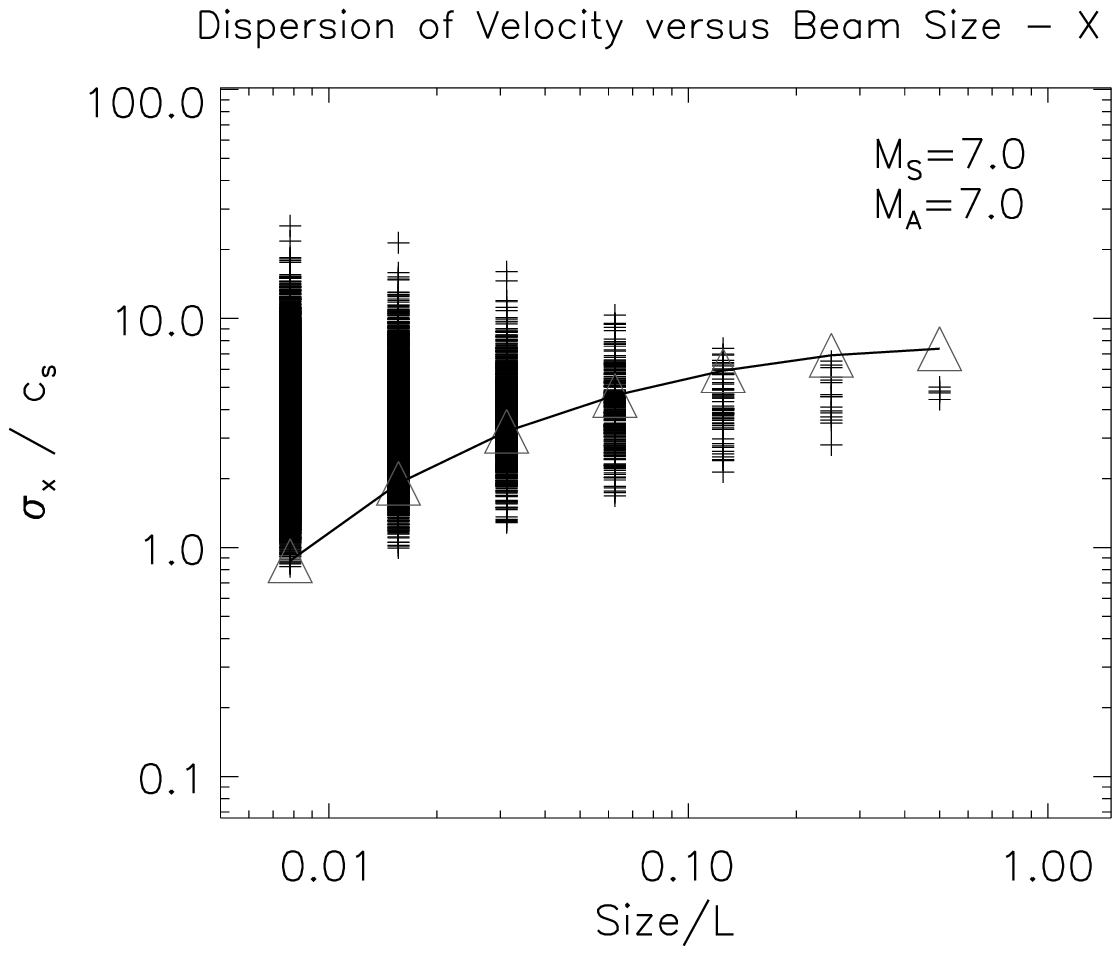}\\

\caption{Dispersion of velocity for models $1-6$ obtained for different beam sizes 
($l^2$) integrated along x-direction. The solid line (and triangles) represent the 
actual 
dispersion, defined as the averaged dispersion measured within all cubes with size $l^3$. 
For a given beamsize, the crosses represent each of the synthetic observed dispersion 
within $l^2L$.}\end{figure*}

To obtain the actual dispersion of velocities as a function of the scale $l$, we 
subdivide the box in $N$ volumes of size $l^3$. Then, we calculate the dispersion of 
$v^*$, normalized by the sound speed $c_s$, as the mean value of the local dispersions 
obtained for each subvolume. For the synthetic observational dispersion, we must firstly 
choose a given line of sight (LOS). Here, for the results shown in Fig.\ 2 we adopted 
x-direction. 
After, we subdivide the orthogonal plane (y-z) in squares of area $l^2$, representing the beamsize. 
Finally, we calculate the dispersion of 
$v^*$ within each of the 
volumes $l^2L$, normalized by the sound speed $c_s$, for different values of $l/L$. The 
results of these calculations for each of the non-viscous models, is shown is Fig. 2. 

The solid line and the triangles represent the average of the actual mean
dispersion of velocities, while the crosses represent each of the synthetic observed 
dispersion within $l^2L$. Regarding the synthetic observational measurements, we see 
that increasing $l$ results in a decrease in the dispersion, i.e. range of values, of 
$\sigma_{v^*_x}$. 
Also, since we use the density weighted velocity $v^*$, the mass distribution plays an 
important role in the calulation of $\sigma_{v^*}$. Denser regions will give a higher 
weight for their own local velocities and, therefore, if several uncorrelated 
denser regions 
are intercepted by the LOS, $\sigma_{v^*}$ will probably be larger. Therefore, we may 
understand the minimum value of $\sigma_{v^*_x}(l)$ as the dispersion obtained for 
the given LOS that intercepts the lowest number of turbulent sub-structures. If a single 
turbulent structure could be observed, then $\sigma_{v^*_x}(l)$ would tend to the actual 
volumetric value if the overdense structure depth is $\sim l$. Also, as you increase $l$, 
the number of different structures 
intercepting the line of sight increases, leading to larger values of the minimum 
observed dispersion. On the other hand, the maximum observed dispersion is directly 
related to the LOS that intercepts most of the different turbulent structures. Since this 
number is unlikely to change, the maximum observed dispersion decreases with $l$ simply 
because of the larger number of points for statistics. As $l \rightarrow L$, 
$\sigma_{v^*_x}(l)$ gets closer to the actual volumetric dispersion. {\bf However, as noted in 
Fig.\ 2, the obtained values for $l \rightarrow L$ are slightly different. This is 
caused by the anisotropy in the velocity field regarding the 
magnetic field, as the velocity components may be different along and perpendicular to 
${\bf B}$. 

Despite of this effect, the results presented in this work do not change when a different orientation for the 
line of sight is chosen. Even though not shown in Fig.\ 2, we have calculated the dispersion of velocity 
for LOS in y and z-directions. 
The general trends shown in Fig. 2 are also observed, but 
a slight difference appears as $l \rightarrow L$, exactly as explained above. This difference is 
expected to be seen in sub-alfvenic cases because of the anisotropy in the velocity distribution.}

It is clear from Fig. 2 that the actual dispersion 
of velocities and a given observational line-width may be very different. Li \& 
Houde (2008), based on Ostriker et al.'s work, assumed that if one chooses, from a 
large number of observational measurements along different LOS's, the minimum 
observational dispersion as the best estimation for the actual dispersion, the associated 
error is minimized. Considering the broad range of observed dispersions 
obtained from the simulations for a given $l$, the minimum value should correspond to the 
actual velocity dispersion. Actually, from Fig. 2 we see that the validity of such 
statement depends on $l$ and on the turbulent regime, though as a general result the 
scaling of the minimum observed dispersion follows the actual one.

For the subsonic models, the actual dispersion is lower than $\sigma_{v^*_x}(l)$, with 
increasing difference as $l/L \rightarrow 0$. In these models, we see that for $l > 0.05 
L$ there is a convergence of the actual dispersion to the synthetic observational 
measurements. At these scales the minimum 
value of $\sigma_{v^*_x}(l)$ is a good estimate of the velocity dispersion of the 
turbulence at the given scale $l$. The difference between both values is less than a 
factor of 3 for all $l$'s, being of a few percent for $l > 0.03 L$. In this turbulent 
regime, mainly at the smaller scales, $\sigma_{v^*_x}(l)$ overestimates the true 
dispersion. For larger scales the associated error is very small and the two quantities 
give similar values.

On the other hand, for the highly supersonic models 
(M$_S \sim 7.0$), the minimum value of 
$\sigma_{v^*_x}(l)$ underestimates the actual value, at most scales. Under this regime, 
the difference to the actual dispersion is of a factor $\sim 2 - 4$. The best matching 
between the two measurements occured for the marginally supersonic cases 
(M$_S \sim 1.5$).

As a major result we found that the uncertainties associated to the NIDR technique 
depend on 
the sonic Mach number of the system, though the associated errors are not extreme in any 
case. We see no major role of the Alfvenic Mach number on this technique.

\subsection{Minimum 2D velocity dispersion $vs$ 3D statistics}

\begin{figure}
\centering
\includegraphics[scale=.6]{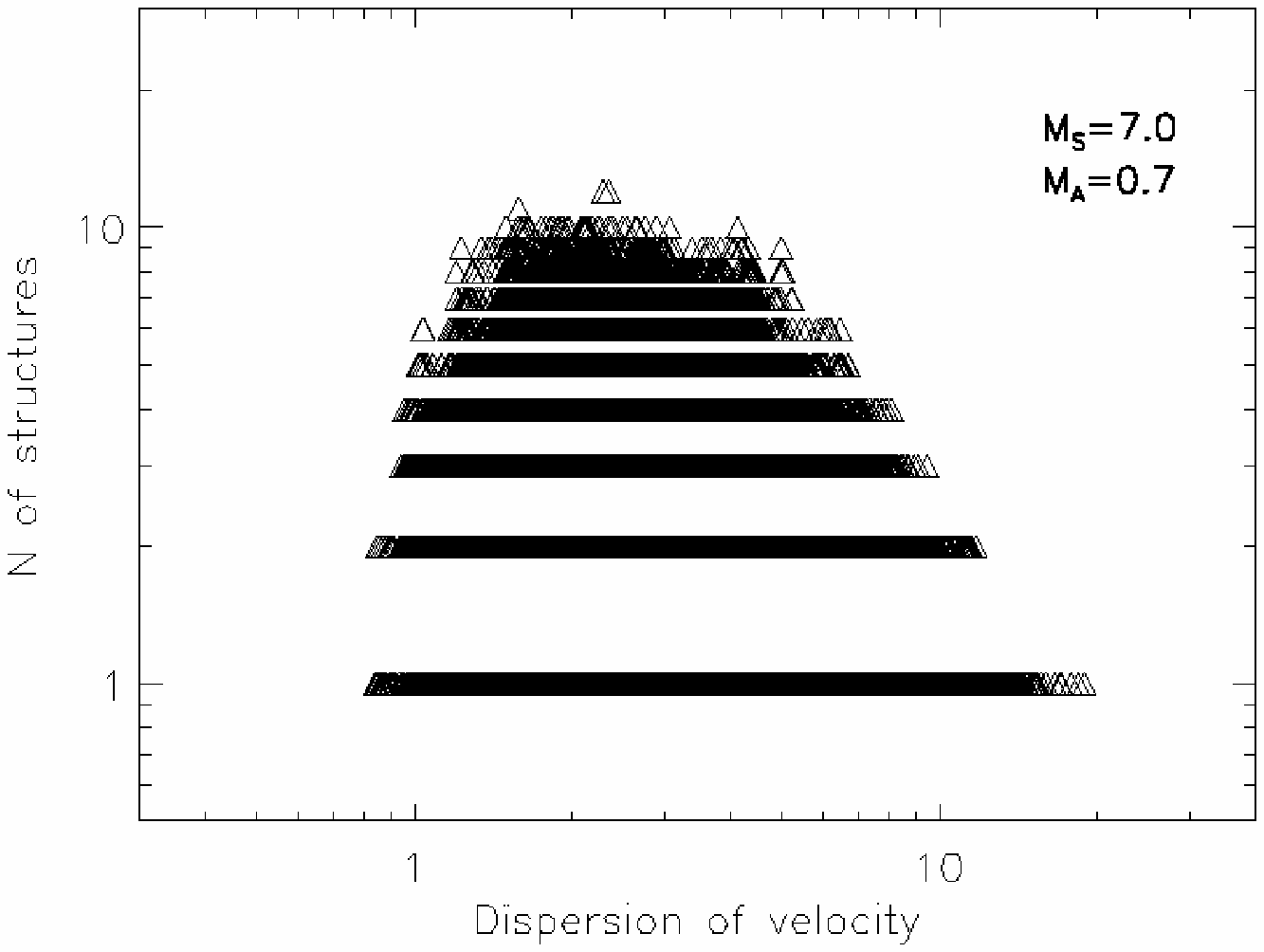}
 
\caption{Correlation of the synthetic velocity dispersion and the number of structures 
intercepted in the line of sight. Each triangle correspond to all LOS's 
calculated from a cube of model 3.}\end{figure}

We have shown that the synthetic observed 
dispersion minima represent a fair approximation for the actual 
3-dimensional dispersion of velocities for the supersonic models, though it is slightly 
overestimated in subsonic cases, and underestimated in highly supersonic cases. What is 
the physical reason for that?

One of the most dramatic differences between subsonic and supersonic turbulence is the 
mass density distribution. Subsonic turbulence is almost incompressible, which means that 
density fluctuations and contrast are small. Supersonic turbulence, on the other hand, 
present strong contrast and large fluctuation of density within the volume. Strong shocks 
play a major role on the formation of high density contrasts, and is the main cause of 
high density sheets and filamentary structures in simulations. Strong shocks also modify 
the turbulent energy cascade, opening the possibility for a more efficient transfer of 
energy from large to small scales, resulting in a steeper 
energy spectra (typically with index $\sim -2$, instead of the Kolmogorov's $\sim -5/3$). 

Observationally, the determination of the velocity dispersion along the line of 
sight is always biased by the density distribution, i.e. a given line profile 
depends on both the emission intensity and the Doppler shifts. For an optically thin 
gas, the emission intensity is directly dependent on the density of the gas. Compared to 
subsonic, we expect the supersonic turbulence to present a larger contrast of density 
in structures. Because of that, the sizes and the number of structures 
intercepting the line of sight may have a deep impact in the determination of the 
observational dispersion of lines. In order to check this hypothesis we calculated the 
number of structures, and their average sizes, along each of the synthetic lines of sight 
of the cube and determined the correlation with the velocity dispersion. {\bf The number 
and depths of the structures were obtained by an algorithm that identifies 
peaks in density distributions. Basically, the algorithm follows three steps. Firstly, 
for each line of sight with beamsize $l^2$, it identifies the maximum peak of density and 
uses a threshold defined as the half value of this maximum of density. Secondly, it 
removes all cells with densities lower than the selected threshold. The remaining data 
represents the dominant structures within the given line of sight.
Finally, the algorithm follows each line of sight detecting the discontinuities, created 
by the use of a threshold, and calculates the sizes of these structures.}

In Fig. 3 we show the correlation between the synthetic velocity dispersion and the 
number of structures intercepted in all LOS, in x-direction, for our cube of the Model 
3. For all models the result is very similar, though not shown in these plots. 
As noticed, the minimum dispersion corresponds to the LOS in which there is only one 
structure intercepted, i.e. there is only one source that dominates the emission line. It 
makes complete sense if this single source is small in depth and, as a volume, we have 
the dispersion of a volume $\sim l^3$. As we increase the number of structures 
intercepted by the LOS, each one contributes with a different Doppler shift, resulting in 
a larger dispersion. However, as we can see from Fig. 3, the maximum dispersion also 
corresponds to a single structure in the LOS. The reason is that we calculate the 
threshold for capturing clumps with the FWHM of the highest peak of the given LOS. If the 
LOS intercepts ``voids'', which are typically very large compared to clumps, the 
algorithm 
results in a number of structures equal {\it one} but its depth, and consequently also 
its velocity dispersion, is large. Taking into account observational sensitivity, 
these low density regions are irrelevant. Furthermore, this picture is also useful to 
understand the scaling relation of $\sigma (l)$. As we increase $l$, increasing the 
beamsize, the number of structures in the LOS is higher resulting in the increase in 
velocity dispersion.

In Fig. 4, we compare the average number of structures intercepting the LOS's and 
their average sizes as a function of the sonic Mach number. It is shown that both the 
number of sources and their intrinsic sizes are inverselly correlated with the sonic Mach 
number. Subsonic models present lower contrast of densities, which corresponds to larger 
overdense structures. For $M_{\rm S} = 0.7$ we obtained a range of densities $0.3 < 
\rho/\bar{\rho} < 3$. As discussed above, large structures correspond to integrated 
volumes $\sim l^2 L$, which deviates from the actual 3D velocity dispersion. The result 
is that the minimum synthetic dispersion obtained from subsonic turbulence will 
overestimate the actual value. For supersonic models, where clumps are systematically 
small, the dispersion minima correspond, from Fig. 3, to the single structures with 
volumes $\sim l^3$, very close to the actual values. On the other hand, for $l 
\rightarrow L$, the result is underestimated. Here, for $M_{\rm S} = 1.5$ we obtained a range of 
densities $0.08 < \rho/\bar{\rho} < 10$, while for $M_{\rm S} = 7.0$ we obtained $0.01 < 
\rho/\bar{\rho} < 90$.

\begin{figure*}
\centering
\includegraphics[scale=1.0]{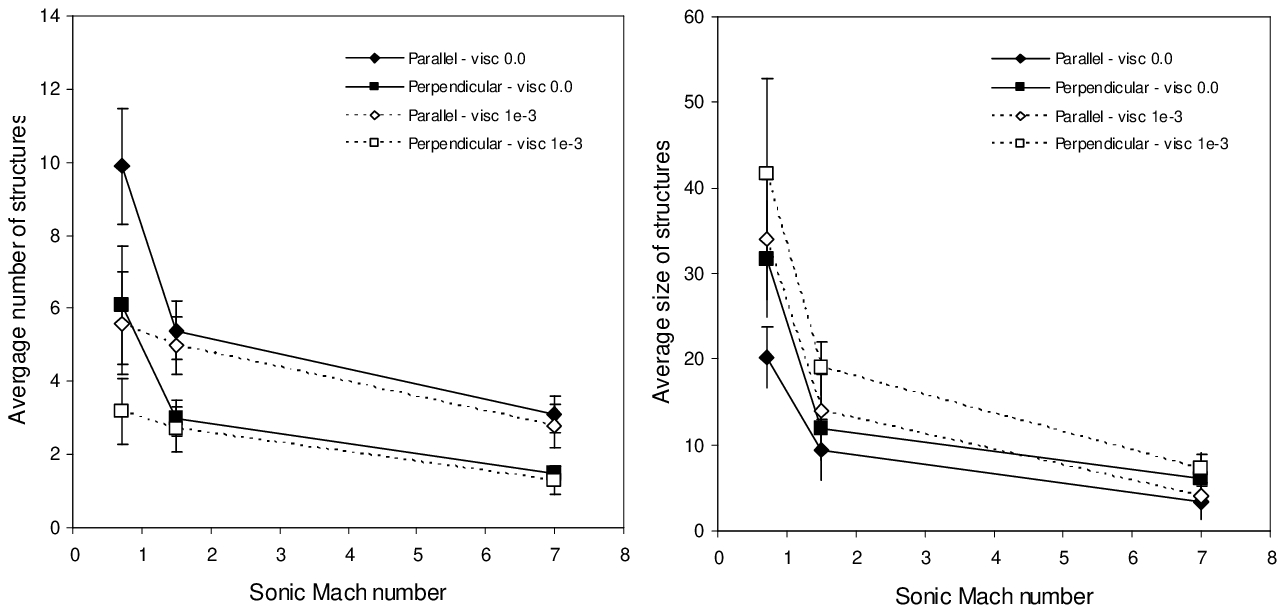}
 
\caption{Average number of structures within all LOS's computed for all models as a 
function of the sonic Mach number.}\end{figure*}

\subsection{Turbulence dissipation scales}

\begin{figure*}[tbh]
\centering
\includegraphics[scale=.60]{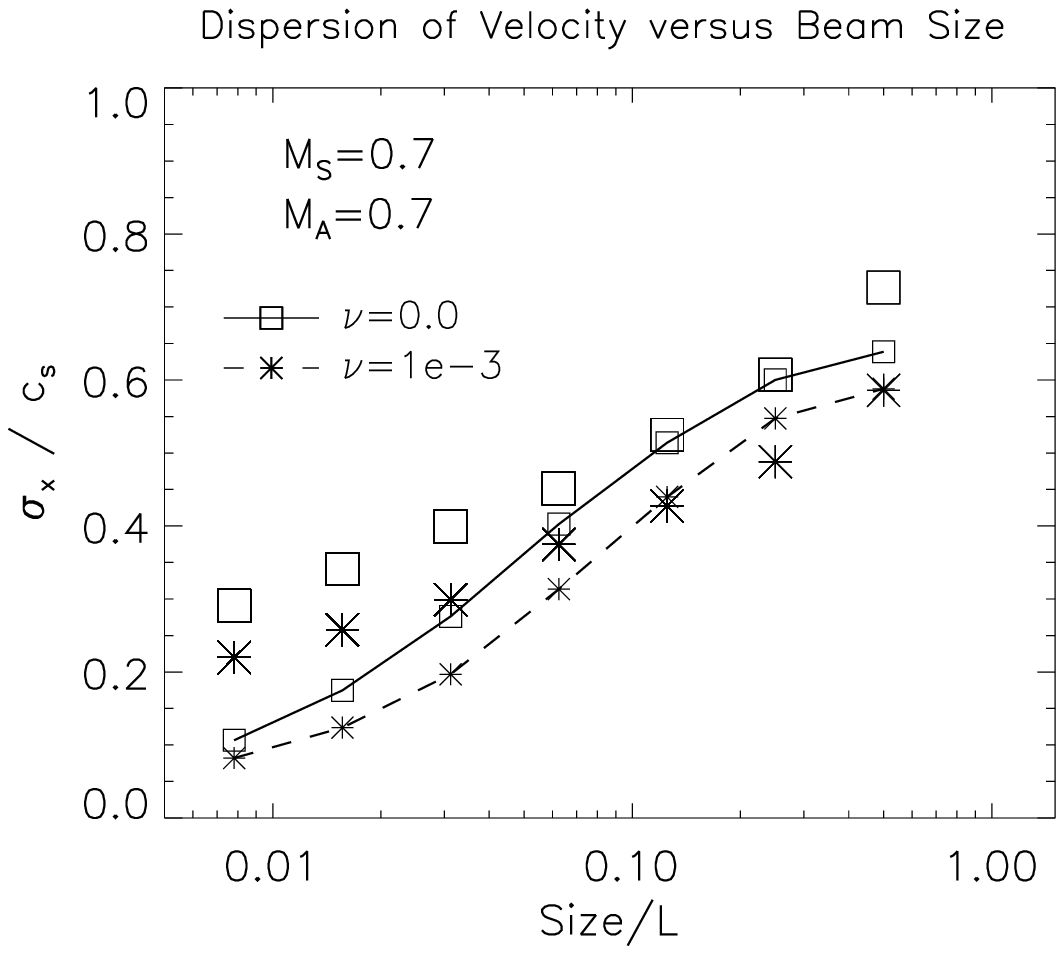}   
\includegraphics[scale=.60]{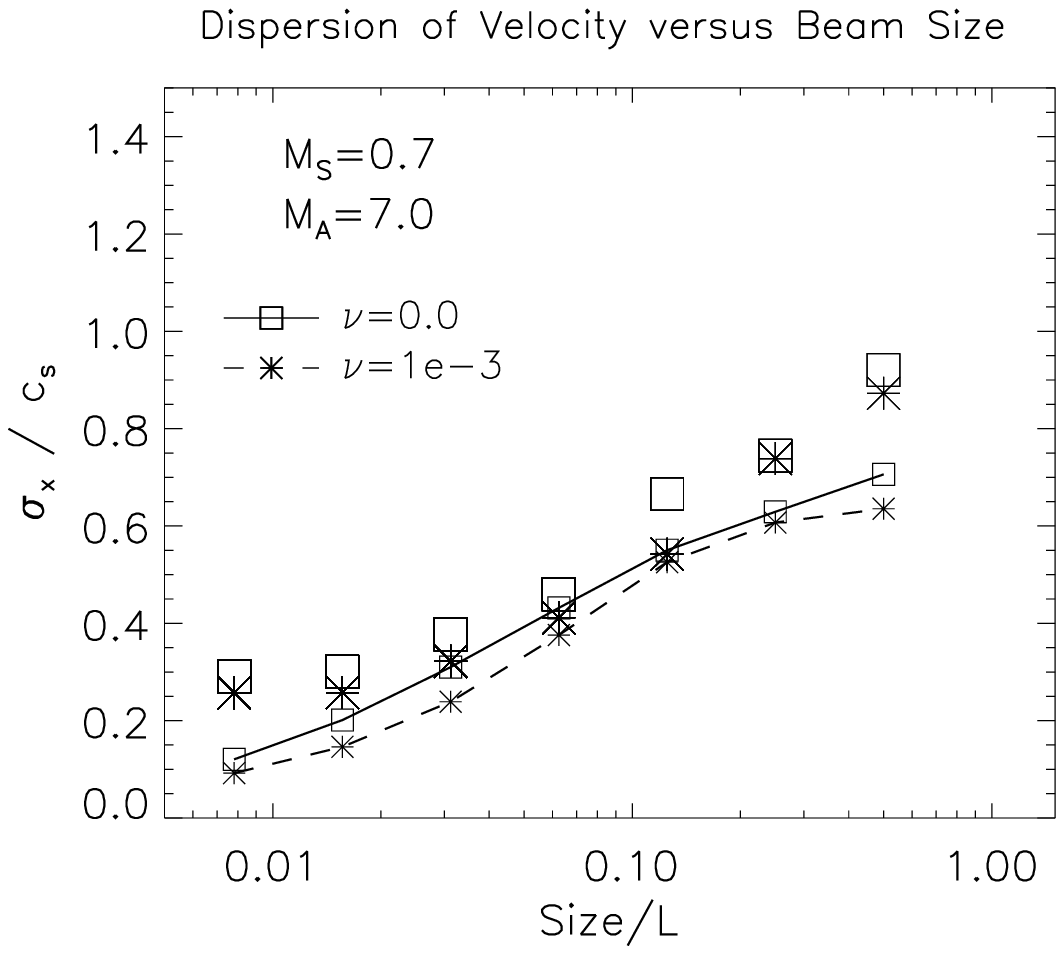}\\ 
\includegraphics[scale=.60]{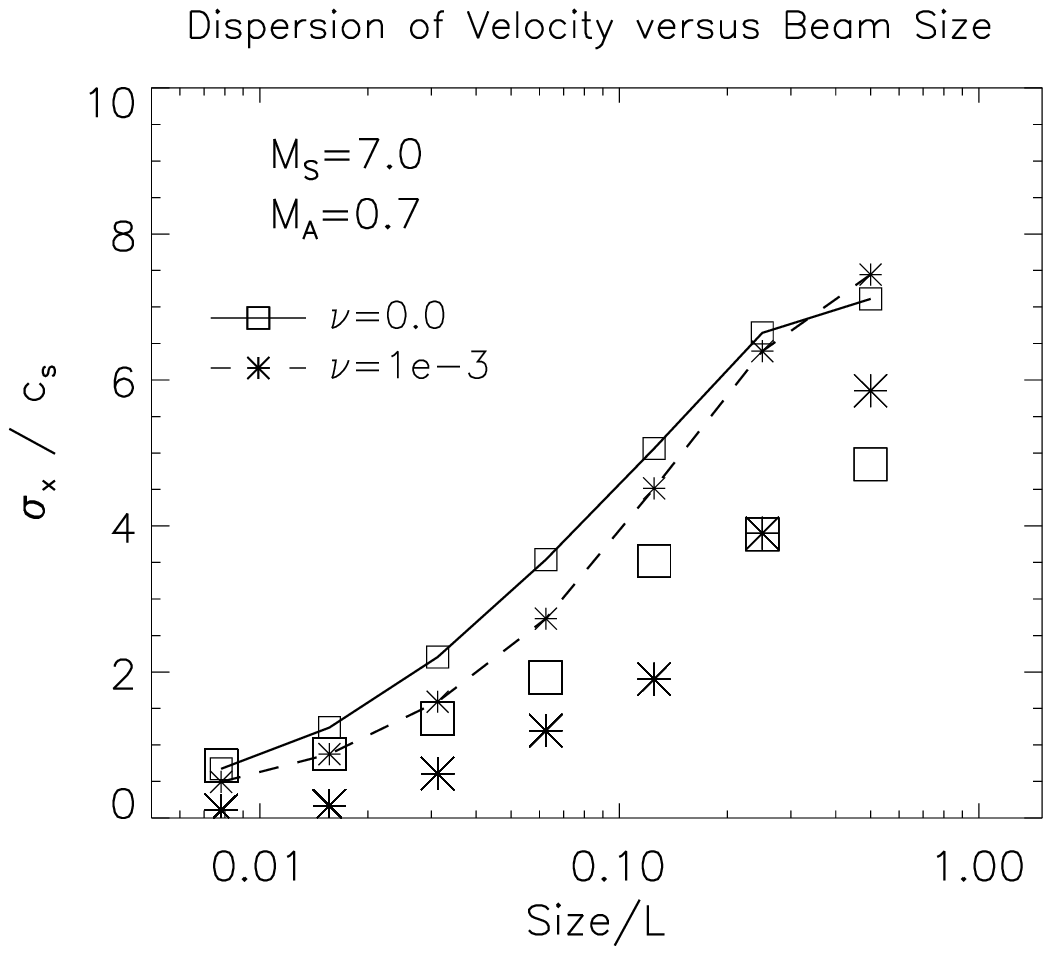}   
\includegraphics[scale=.60]{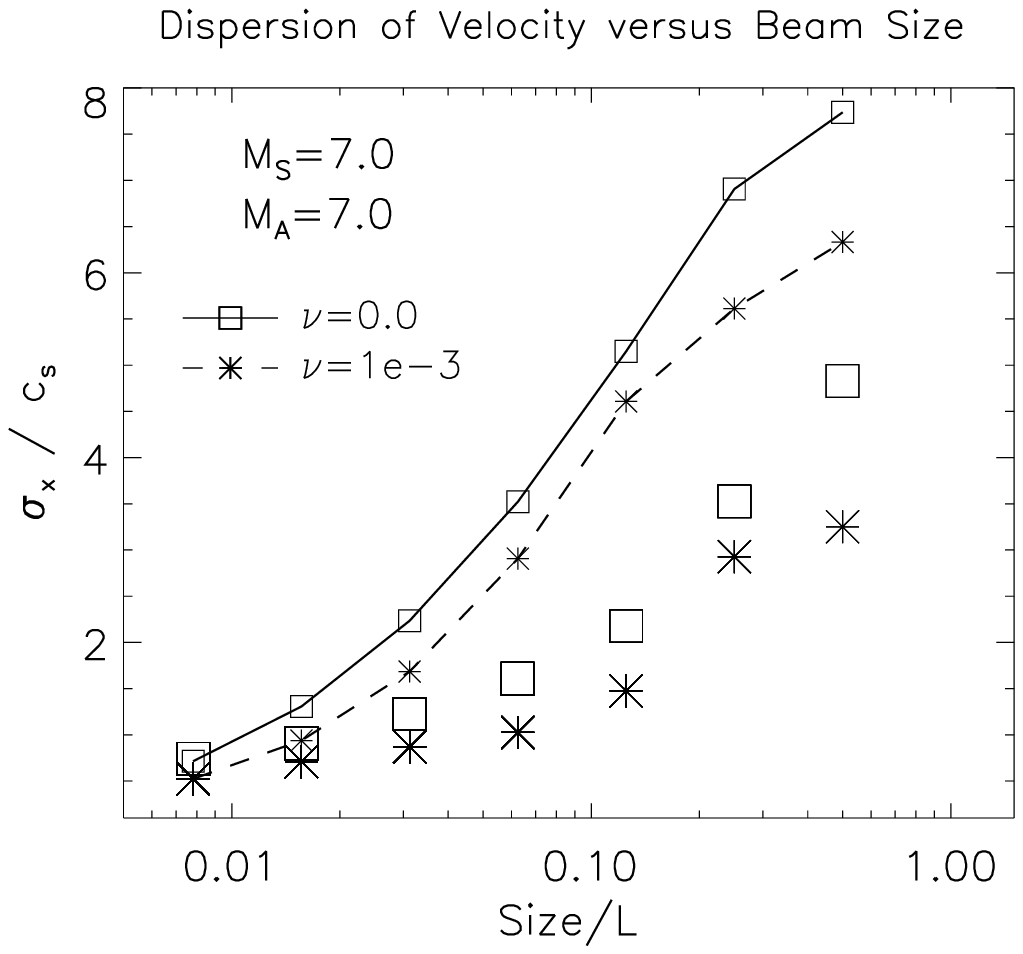}\\  

\caption{Actual (lines) and the minimum observed (symbols) dispersions of velocity 
 for the viscous (stars) and inviscid fluids (squares).}
\end{figure*}

\begin{table*}
\begin{center}
\caption{Parameters of best fit and damping scales}
\begin{tabular}{ccccccccccc}
\hline\hline
 & & &  & actual &  & &  & synthetic maps &  &  \\
$M_S$ & $M_A$ & & b$_{\rm ac}$ & n$_{\rm ac}$ & $L_D^{\rm ac}$ & & b$_{\rm obs}$ & 
n$_{\rm obs}$ & $L_D^{\rm obs}$ & $L_D^{\rm ac} / L_D^{\rm obs}$ \\\hline
$0.7$ & $0.7$ & & 0.63(16) & 0.39(2) & 0.244(21) & & 1.01(9) & 0.80(5) & 0.088(13) & 2.7 \\
$0.7$ & $7.0$ & & 1.38(12) & 0.65(3) & 0.246(21) & & 1.00(10) & 0.76(4) & 0.215(17) & 1.1 \\
$7.0$ & $0.7$ & & 1.58(21) & 1.01(4) & 0.162(17) & & 1.58(13) & 1.08(5) & 0.318(23) & 0.5 \\
$7.0$ & $7.0$ & & 1.82(18) & 1.05(4) & 0.041(6) & & 2.51(18) & 0.95(6) & 0.197(25) & 0.2 \\
\hline\hline
\end{tabular}
\tablenotetext{}{$L_D^{\rm ac}$ and $L_D^{\rm obs}$ are placed in terms of the total size of the box, i.e. $L_D/L$.}
\end{center}
\end{table*}

In order to obtain the dissipation scales of the ionized flows, we must apply Eqs. (5) 
and (6) and 
2 to the simulated data. For the inviscid flows, we calculate $b$ and $n$ (Eq.5) using 
both the actual and minimum observed velocity dispersions. Then, we use the data from the 
viscous simulation to calculate the difference $\sigma^2_{\rm n} - \sigma^2_{\rm i}$. 
Finally, from Eq.(6), knowing $\sigma^2_{\rm n} - \sigma^2_{\rm i}$, $n$ and $b$ it is 
possible to obtain $k_{\rm D}$. In Fig. 5 we show the data used to compute Eqs.(5) and 
(6), 
i.e. both the actual (lines) and the minimum observed (symbols) dispersions of velocity 
 for the viscous (stars) and inviscid fluids (squares). The fit parameters (Eq.5) for 
the inviscid simulated data are shown in Table 2, where b$_{\rm ac}$ and n$_{\rm ac}$ 
where obtained for the actual velocity dispersion and b$_{\rm obs}$ and n$_{\rm obs}$ for 
the synthetic observed line widths. We see that $n$ increases with the sonic Mach 
number (M$_S$), as explained below. In Table 2, we also present 
the damping scale $L_{D}$, obtained from Eq.(6). In the last column we show the ratio 
between the actual and observational scales $L_D^{\rm ac} / L_D^{\rm obs}$. The ratio of 
the obtained scales showed a maximum difference of a factor of 5 between the actual value 
of the dispersion and the one obtained from the synthetic observational maps. Also, 
compared to the expected values obtained visually from the spectra (Fig.1), there is a 
good correspondence with $L_{D}^{\rm obs}$ given in Table 2. This fact shows that the 
method indeed might be useful. 

Regarding the spectral index $\alpha$ (Eq.1), Table 2 gives $\alpha \sim 1.4 - 1.8$ for 
the subsonic models, and $\alpha \sim 2.0 - 2.1$ for the supersonic models. These 
parameters may be directly compared to the values $\alpha \sim 1.7$ and $\alpha \sim 2.0$ 
expected for theoretical incompressible and compressible turbulent spectra, respectively. 
Also, the increase of $b$ as $M_S$ increases shows that the energy transfer rate is 
larger for compressible models. From Table 2, we obtain an averaged value of 
$\dot{\epsilon} \sim 0.6$ for subsonic and 1.4 for supersonic models, in code units.

\subsection{Accuracy of the method}

Before describing a potential technique for future observations, we must stress out that 
this work may be divided in two independent parts. The first is related to the 
characterization of the energy power spectrum of turbulence, including the energy 
transfer rate and cut-off 
length, while the second is based on the use of this damping length which is used in the Li \& Houde (2008) approach to determine the magnetic field.

As 
mentioned before, in magnetized partially ionized gases, several different mechanisms may 
act in order to damp turbulent motions. The determination of the magnetic field intensity 
from the damping scale is strictly dependent on the assumption that the turbulent cut-off 
is due to the ambipolar diffusion, i.e. the ion-neutral diffusion scale is larger than 
the scales of any other dissipation mechanism. Unfortunately, this could not be tested in 
the simulations since we did not include explicit two fluid equations to check the 
role of ambipolar diffusion and other damping mechanisms in the turbulent spectra.

In the simulations, the cut-off is obtained via an explicit and arbitrary viscous 
coefficient. The result is clear in the power spectra (Fig. 1), where the damping length 
shifts to lower values of $k$. From those, we found out that the ``observational" 
velocity dispersions for a given beam size $l \times l$, in most cases, do not coincide 
with the actual dispersion at scale $l$. Also, the estimation from the minimum value of 
the observed dispersions may also be different from the expected measure. The associated 
errors 
depend on the sonic Mach number and on the scale itself, as shown in the previous 
section. However, the parameters 
obtained for the power-law of the synthetic observations and actual 3-dimensional 
distributions showed to be quite similar. 
The theory behind this method is very simple, but still reasonable, as 
it assumes that the two fluids (ions and neutrals) would present the same 
cascade down to the dissipation scale, when they decouple, where the ion turbulence 
would be sharply damped. We believe that these errors may, eventually, be originated by 
the short inertial range of the simulated data. We see from Fig. 1 that the turbulent 
damping range is broad, and not sharp compared to the inertial range, as assumed in this 
model. In the real ISM, the power spectrum presents a constant slope within a much 
broader range, and the errors with real data may be smaller.

\section{Discussion}

In this work we studied the relationship between the actual 3-D 
dispersion of velocities and the ones obtained from synthetic observational line 
profiles, i.e. the density weighted line profile widths, along 
different LOS's. We study the possibility of the scaling relation of the turbulent 
velocity dispersion being determined from spectral line profiles. 
If correct, the observed line profiles could allow us to determine in 
details the turbulence cascade and the dissipation lengths of turbulent eddies in the 
ISM. However, in order to check the validity of this approach, we performed a number of 
higher resolution turbulent MHD simulations under different turbulent regimes, i.e. for 
different sonic and Alfvenic Mach numbers, and for different viscosity coefficients.

Based on the simulations, we showed that the synthetic observed line width 
($\sigma_{\rm v}$) is related to the number  
of dense structures intercepted by the LOS. Therefore, the actual dispersion at scale $l$ 
tends to be similar to the line width obtained by a LOS within a beam size $l \times l$ 
intercepting a single dense structure. It means that, the minimum 
observed dispersion may be the best estimative for the actual dispersion of velocities, 
if a large number of LOS is considered, as assumed in the theoretical model.

Moreover, by adopting a power-law for the spectrum function of $\sigma_{\rm v}$, we were 
able to estimate the spectral index $n$ and constant $b$ (Eq. 5), given in Table 2. We 
see a good correspondence between the parameters obtained for the 3-D dispersions and 
synthetic line profiles. Furthermore, since $n$ is associated with the turbulent spectral 
index $\alpha$, and $b$ with the energy tranfer rate between scales, line profiles may be 
useful in characterizing the turbulent cascade.

Also, we showed that the models under similar turbulent regime but with different 
viscosities will result in different dispersions of velocity, on both 3-D and synthetic 
line profile measurements. The difference of $\sigma_{\rm v}$ for the inviscid and 
viscous models, associated with the parameters $n$ and $b$ previously obtained, gives an 
estimative of the damping scale $L_{\rm D}$ of the viscous model (Eq. 6). The ratio 
between the damping scales obtained from the 3-D and synthetic profile dispersions vary 
only by a factor $\leq 5$.

A good estimate of the cut-off scale of the ISM turbulence may bring light to much 
of the uncertainties about the mechanisms that are responsible for the damping of 
the turbulent eddies. As we showed, if ambipolar diffusion is the main 
phenomenon responsible 
for the dissipation of turbulence, then it is possible to provide another method for 
the determination of magnetic field strength in dense cores, besides Zeeman splitting.

We believe that, i - the numerical resolution used in our models, and ii - the single 
fluid aproximation, with different viscosities to simulate neutrals and ions separately, 
are the main limitations of this work. The ISM may present more than 5 decades of 
inertial range in its power spectrum while numerical simulations are, at best, 
limited to 2-3. On the other hand, since the validity of the theoretical approximation 
presented in this work depends on the broadness of the inertial range, we expect this 
model to be even more accurate for finer resolution, or for the ISM itself. Under the 
single 
fluid aproximation made in this work we fail in correlating the ambipolar diffusion with 
increased viscosity and, therefore it is not possible to test the estimation of $B$ from 
the damping scales as proposed above, though the results related to the damping length 
and energy transfer rates remain unchanged. In a two-fluid simulation this is more likely 
to be fulfilled. We are currently implementing the two-fluid set of equations 
in the code, and will test this hypothesis in the future.

\section{Summary}

In this work we presented an extensive analysis of the applicability of the 
NIDR method for the determination of the turbulence damping scales and the magnetic 
field intensity, if ambipolar diffusion is present, based on numerical
simulations of viscous MHD turbulence. We performed simulations with 
different characteristic sonic and Alfvenic Mach numbers, and different explicit viscous 
coefficients to account for the physical damping mechanisms. As main results we showed 
that:

\begin{itemize}
\item the correspondence between the synthetic observational dispersion of 
velocities (i.e. from the 2D oserved maps) and the actual 3-dimensional dispersion of 
velocities depends on the  
turbulent regime;

\item for subsonic turbulence, the minimum inferred dispersion tends to overestimate the 
actual dispersion of velocities for small scales ($l << L$), but presented good 
convergence at large scales ($l \rightarrow L$);

\item for supersonic turbulence, on the other hand, there is a convergence at small 
scales ($l << L$), but the minimum inferred dispersion tends to underestimate the 
actual dispersion of velocities at large 
scales ($l \rightarrow L$);

\item even though not precisely matching, the actual velocity and 
the minimum velocity dispersion from spectral lines were well fitted by a power-law 
distribution. We 
obtained similar slopes and linear coefficients for both measurements, with $\alpha \sim 
-1.7$ and 
$-2.0$ for subsonic and supersonic cases, respectivelly, as expected theoretically;

\item the damping scales obtained from the fit for the both cases were similar. 
The difference between the scales obtained from the two fits was less than a factor of 5 for 
all models, indicating that the method may be robust and used for observational data;
 
\end{itemize}

The work presented in this paper tests some of the key assumptions important process by technique of
Li \& Houde (2008). Evidently, more work is still required in order to test the full range of applicability of the method (e.g., test 
cases of both magnetically and neutral driven turbulence). The aforementioned implementation of two-fluid 
numerical simulations to better mimic ambipolar diffusion will be an important step in that 
direction.

\acknowledgments

D.F.G. thank the financial support of the Brazilian agencies FAPESP (No.\ 
2009/10102-0) and CNPq (470159/2008-1), and the Center for Magnetic Self-Organization in Astrophysical and Laboratory Plasmas (CMSO). A.L. acknowledges NSF grant AST 0808118 and the CMSO. M.H.'s research is funded through the NSERC Discovery Grant, Canada Research Chair, Canada 
Foundation for Innovation, Ontario Innovation Trust, and Western's Academic Development 
Fund programs. The authors also thank the anonymous referee whose useful comments and 
suggestions helped to improve the paper.


\begin{thebibliography}{}

\bibitem[]{} Beresnyak, A. \& Lazarian, A. 2006, \apj, 640, 175

\bibitem[]{} Beresnyak, A. \& Lazarian, A. 2009, \apj, 702, 460

\bibitem[]{} Biskamp, D. 2003, {\it Magnetohydrodynamic Turbulence}, Cambridge University 
Press.

\bibitem[]{} Boldyrev, S. 2005, \apj, 626, 37

\bibitem[]{} Boldyrev, S. 2006, Physical Review Letters, 96, 5002

\bibitem[Brunt \& Heyer (2002)]{brunt02}
  Brunt, C. M., \& Heyer, M. H. 2002, \apj, 566, 289

\bibitem[Burkhart et al.(2009)]{burk09}
  Burkhart, B., Falceta-Gon\c calves, D., Kowal, G. \& Lazarian 2009, \apj, 693, 250

\bibitem[Cho \& Lazarian(2002)]{cho02} Cho, J. \& Lazarian, A. 2005, Physical Review 
Letters, 88, 5001

\bibitem[Cho \& Lazarian(2003)]{cho03} Cho, J. \& Lazarian, A. 2005, \mnras, 345, 325

\bibitem[Cho \& Lazarian(2005)]{cho05} Cho, J. \& Lazarian, A. 2005, Theor. Comput. Fluid 
Dynamics, 19, 127

\bibitem[Crutcher et al.(1999)]{cru99b} Crutcher, R. M., Roberts, D. A., Troland, T. H. 
\& Goss, W. M. 1999, \apj, 515, 275

\bibitem[Crutcher(2004)]{crutcher04}
  Crutcher, R. 2004, Ap\&SS, 292, 225

\bibitem[Del Zanna, Bucciantini \& Londrillo(2003)]{delzanna03}
  Del Zanna, L., Bucciantini, N. \& Londrillo, P.  2003, \aap, 400, 397

\bibitem[Draine \& Lazarian(1999)]{dl99} Draine, B. T. \& Lazarian, A. 1999, ApJ, 512, 740

\bibitem[Elmegreen \& Scalo(2004)]{elm04}
  Elmegreen, B. \& Scalo, J. 2004, \araa, 42, 211

\bibitem[Esquivel \& Lazarian(2005)]{el05}
Esquivel, A. \& Lazarian, A. 2005, \apj, 631, 320

\bibitem[Falceta-Gon\c calves, Lazarian \& Kowal(2008)]{falceta08} Falceta-Gon\c calves, 
D., Lazarian, A. \& Kowal, G. 2008, \apj, 679, 537

\bibitem[Falceta-Gon\c calves et al.(2010)]{falceta10} Falceta-Gon\c calves, 
D., de Gouveia Dal Pino, E. M., Gallagher, J. S. \& Lazarian, A. 2010, \apj, 708, 57

\bibitem{falc03} Falceta-Gon\c calves, D., de Juli, M. C. \& Jatenco-Pereira, V. 2003, 
\apj, 597, 970

\bibitem[Fiedge \& Pudritz(2000)]{fiedge00} Fiedge, J. D. \& Pudritz, R. E. 2000, \apj, 
544, 830

\bibitem[Goldreich \& Sridhar(1995)]{gs95} Goldreich, P. \& Sridhar, S. 1995, \apj, 438, 
763 

\bibitem[Harten, Lax \& van Leer(1983)]{hart83} Harten, A., Lax, P. D. \& van Leer, B. 1983, SIAM Rev., 25, 35

\bibitem[]{} Higdon, J. C. 1984, \apj, 285, 109

\bibitem[Hildebrand et al.(2000)]{hil00} Hildebrand, R. H., Davidson, J. A., Dotson, J. 
L., Dowell, C. D., Novak, G., \& Vaillancourt,J. E. 2000, \pasp, 112, 1215

\bibitem[Hildebrand et al.(2009)]{hil09} Hildebrand, R. H., Kirby, L., Dotson, J. L, 
Houde, M.,\& Vaillancourt,J. E. 2009, \apj, 696, 567

\bibitem[Houde et al.(2000a)]{hou00a} Houde, M., Bastien, P., Peng, R., Phillips, T.G. \& 
Yoshida, H. 2000a, \apj, 536, 857

\bibitem[Houde et al.(2000b)]{hou00b} Houde, M., Peng, R., Phillips, T.G.,  Bastien, P. 
\& Yoshida, H. 2000b, \apj, 537, 245


\bibitem[Houde et al.(2009)]{hou09} Houde, M., Vaillancourt,J. E., Hildebrand, R. H., Chitsazzadeh, S., \& Kirby, L. 2009, \apj, in press

\bibitem{iri64} Iroshnikov, P. 1964, Sov. Astron., 7, 566

\bibitem[Kowal, 
Lazarian \& Beresniak(2007)]{kowal07} Kowal, G., Lazarian, A. \& 
Beresniak, A. 2007, \apj, 658, 423

\bibitem[Kowal \&
Lazarian(2007)]{kowal07b} Kowal, G. \& Lazarian, A. 2007, \apj, 666, 69

\bibitem[Kowal et al.(2009)]{kowal09} Kowal, G., Lazarian, A., 
Vishniac, E.~T., \& Otmianowska-Mazur, K.\ 2009, \apj, 700, 63

\bibitem{kra65} Kraichnan, R. 1965, Phys. Fluids, 8, 1385

\bibitem[Kritsuk et al.(2007)]{kritsuk07} Kritsuk, A. G., Norman, M. L., Padoan, P. \& 
Wagner, R. 2007, \apj, 665, 416

\bibitem[]{} Lazarian, A. \& Vishniac, E. T. 1999, \apj, 517, 700

\bibitem{laz04} Lazarian, A., Vishniac, E. T. \& Cho, J. 2004, 
\apj, 603, 180

\bibitem{laz05} Lazarian, A. 2005, Bulletin of the American Astronomical Society, 37, 1358

\bibitem[]{} Lazarian, A. 2007, Journal of Quantitative Spectroscopy \& Radiative 
Transfer, 106, 225

\bibitem[Le{\~a}o et al.(2009)]{leao09} Le{\~a}o, M.~R.~M., de 
Gouveia Dal Pino, E.~M., Falceta-Gon{\c c}alves, D., Melioli, C., 
\& Geraissate, F.~G.\ 2009, \mnras, 394, 157

\bibitem[Li \& Houde(2008)]{li08} Li, H. \& Houde, M. 2008, \apj, 677, 1151

\bibitem[]{} Lithwick, Y. \& Goldreich, P. 2001, \apj, 562, 279

\bibitem[Londrillo \& Del Zanna(2000)]{londrillo00} Londrillo, P. \& Del Zanna, L. 2000, \apj, 530, 508

\bibitem[MacLow \& Klessen(2004)]{maclow04} MacLow, M. M. \& Klessen, R. S. 2004, Rev. 
Mod. Phys., 76, 125

\bibitem[]{} McKee, C. F. \& Ostriker, E. C. 2007, \araa, 45, 565

\bibitem[]{} Mestel, L. \& Spitzer, L., Jr. 1956, \mnras, 116, 503

\bibitem{mint97} Minter, A. H. \& Spangler, S. R. 1997, \apj, 458, 194

\bibitem[Ostriker, Stone \& Gammie(2001)]{osg01} Ostriker, E. C., Stone, J. M. \& Gammie, 
C. F. 2001, \apj, 546, 980

\bibitem[]{} Santos de Lima et al. 2010, \apj, accepted

\bibitem[]{} Shebalin, J. V., Matthaeus, W. H. \& Montgomery, D. 1983, Journal of Plasma 
Physics, 29, 525

\bibitem[]{} Shu, F. H. 1983, \apj, 273, 202

\bibitem[]{} Yan, H. \& Lazarian, A. 2006, \apj, 653, 1292

\bibitem[]{} Yan, H. \& Lazarian, A. 2007, \apj, 657, 618

\bibitem[]{} Yan, H. \& Lazarian, A. 2008, \apj, 677, 1401

\bibitem[]{} Zank, G. P. \& Matthaeus, W. H. 1992, \jgr, 97, 17, 189


\end{thebibliography}
\end{document}